\documentclass[aps,twocolumn,a4paper]{revtex4-1}

\usepackage{graphicx}
\usepackage{amsfonts}

\usepackage{amssymb}
\usepackage{amsmath} 

\usepackage{multirow} 
\usepackage{tabularx} 
\usepackage{array}

\usepackage{braket}

\usepackage{color}
\usepackage{bm}
\usepackage{hyperref}
\usepackage{comment}

\renewcommand{\Im}{\operatorname{{\mathrm Im}}}

\allowdisplaybreaks[3]

\begin{document}

\title{Microscopic theory of magnon-drag electron flow in ferromagnetic metals} 
\author{Terufumi Yamaguchi }
\affiliation{Department of Physics, Nagoya University, Nagoya 464-8602, Japan} 
\author{Hiroshi Kohno}
\affiliation{Department of Physics, Nagoya University, Nagoya 464-8602, Japan} 
\author{Rembert A. Duine}
\affiliation{Institute for Theoretical Physics, Utrecht
University, Leuvenlaan 4, 3584 CE Utrecht, The Netherlands}
\affiliation{Department of Applied Physics, Eindhoven University of Technology,
P.O. Box 513, 5600 MB Eindhoven, The Netherlands}
\date{\today}

\begin{abstract}
 A temperature gradient applied to a ferromagnetic metal induces not only independent 
flows of electrons and magnons but also drag currents because of their mutual interaction. 
 In this paper, we present a microscopic study of the electron flow induced by the drag due to  
magnons. 
 The analysis is based on the \textit{s-d} model, which describes conduction electrons 
and magnons coupled via the \textit{s-d} exchange interaction. 
 Magnetic impurities are introduced in the electron subsystem as a source of spin relaxation. 
 The obtained magnon-drag electron current is proportional to the entropy of magnons 
and to $\alpha - \beta$ (more precisely, to $1 - \beta/\alpha$), 
where $\alpha$ is the Gilbert damping constant and 
$\beta$ is the dissipative spin-transfer torque parameter. 
 This result almost coincides with the  previous phenomenological result based on 
the magnonic spin-motive forces, and consists of spin-transfer and 
momentum-transfer contributions, but with a slight disagreement in the former. 
 The result is interpreted in terms of the nonequilibrium spin chemical potential generated by nonequilibrium magnons. 
\end{abstract}

\maketitle

\section{Introduction}

 Transport phenomena in ferromagnetic metals exhibit surprisingly rich physics
as unveiled by intensive studies in spintronics. 
 This is largely because they involve transport of not only charge and heat  
but also spin angular momentum. 
 In the presence of magnetization textures, applying an electric current 
induces magnetization dynamics because of spin-transfer torques 
that the spin current of electrons exerts on the magnetization \cite{torque1,torque2}. 
 In turn, a time-dependent magnetization induces spin and charge currents of electrons 
via spin-motive forces that are reciprocal to the spin-transfer torques \cite{smf}. 
 Even when the (equilibrium) magnetization is uniform, its thermal/quantum fluctuations, 
i.e., spin waves or magnons, can interact with electrons. 
  Moreover, transport through an inhomogeneous region induces nonequilibrium 
spin accumulation, both in electrons and magnons, 
which then induce diffusion spin currents. 
 The concept of \lq\lq spin chemical potential'' \cite{mu_s} 
and \lq\lq magnon chemical potential'' \cite{mu_mag}
have been introduced to describe such effects.

 One of the important effects in the interplay of electrons and magnons in transport phenomena are drag effects. 
 When subjected to a temperature gradient, 
electrons and magnons start to flow, first independently, and then by dragging with each other. 
 Thermoelectric measurements indicate the presence of magnon-drag contributions in 
Fe \cite{Blatt1967}, NiCu \cite{Grannemann1976}, NiFe \cite{Costache2011}, 
and in Fe, Co and Ni  \cite{Watzman2016}.
 Theoretical studies include both phenomenological 
\cite{Grannemann1976, Lucassen2011, Flebus2016} 
and microscopic \cite{Miura2012} ones. 
 In particular, phenomenological studies based on the spin-motive force picture 
\cite{Lucassen2011, Flebus2016} 
indicate the importance of the dissipative $\beta$ parameter, 
which stems from spin relaxation of electrons. 
 Microscopic treatment of spin-relaxation effects requires the consideration 
of so-called vertex corrections, beyond the simple self-energy (damping or scattering-time) 
effects, as noted in the study of current-induced spin torque \cite{Kohno2006,Duine2007}, 
but such studies are not available yet for the drag effects. 
 In a related work, which studies spin torques due to magnons, 
a careful treatment of the spin-relaxation effects revealed an additional contribution 
not obtained in a phenomenological analysis \cite{Yamaguchi2017}. 
 Therefore, one may expect an analogous situation also in magnon-drag transport phenomena.

 In this paper, we present a microscopic analysis of magnon-drag electric current 
(or electron flow) induced by a temperature gradient. 
 Using as a microscopic model the $s$-$d$ model that describes conduction electrons 
interacting with magnons, 
we calculate the electric current caused by magnons that are driven by the temperature gradient. 
 The temperature gradient is treated by its mechanical equivalent, 
a fictitious gravitational field, introduced by Luttinger \cite{Luttinger1964}. 
 The obtained result consists of two terms, which may be interpreted as due to the 
spin-transfer effect and the momentum-transfer effect, as in the phenomenological 
theory \cite{Flebus2016}. 
 However, as to the former (spin-transfer effect), there is a quantitative difference, 
and our result is proportional to $\alpha - \beta$ (or $1 - \beta/\alpha$), 
where $\alpha$ is the Gilbert damping constant. 
 It vanishes, and changes sign, at $\alpha = \beta$, 
which agrees with the intuitive notion that the case $\alpha = \beta$ is very special. 
 Although this is mostly of conceptual importance, it may acquire a practical one 
if one can determine the value of $\beta/\alpha$ from magnon-drag experiments. 
 We interpret the results in terms of the spin chemical potential induced by magnons. 
 In the course of our study, we give an argument that justifies the Luttinger's argument 
by an explicit calculation.

 The organization of the paper is as follows. 
 In Sec.~\ref{sec:model}, we describe the microscopic model and some calculational tools 
such as Green's functions. 
 In Sec.~\ref{sec:calculation},
we outline the microscopic calculation of magnon-drag electron flow. 
 The result is discussed in terms of spin-motive force and spin chemical potential. 
 In Sec.~IV, we revisit the phenomenological theory based on the spin-motive force, 
and compare the result with our microscopic result. 
 In Sec.~V, we give an alternative analysis which \lq\lq derives'' the spin chemical potential. 
 Details of the microscopic calculations are presented 
in Appendices \ref{sec:vertexcorrection} and \ref{sec:micro_cal}. 
 In Appendix \ref{sec:pheno_anal}, we reanalyze the phenomenological theory 
in another way using the stochastic Landau-Lifshitz-Gilbert equation.

\section{Model}\label{sec:model}

\subsection{Hamiltonian}

We consider a system consisting of conduction electrons and magnons 
in a ferromagnetic metal with uniform equilibrium magnetization. 
 The Hamiltonian is given by 
\begin{align}
	\mathcal{H}
	&= 
	\mathcal{H}^{0}_{\rm el} + \mathcal{H}_{{\rm mag}} + \mathcal{H}_{sd},
	\label{eq:toatal_Hamiltonian}
	\\
	& \mathcal{H}^{0}_{{\rm el}}
	= 
	\int {\rm d} \bm{r}
	\left[ 
		\frac{1}{2m} (\partial_{i} c^{\dagger}) (\partial_{i} c) 
		+ c^{\dagger} (V_{\rm imp} - \mu) c \, 
	\right],
	\label{eq:Hamiltonian_el0}
	\\
	& \mathcal{H}_{\rm mag}
	= 
	\sum_{\bm{q}} \omega_{\bm{q}} a^{\dagger}_{\bm{q}} a_{\bm{q}},
	\label{eq:Hamiltonian_mag}
	\\
	& \mathcal{H}_{sd}
	= - J_{sd} \int {\rm d} \bm{r} \, c^{\dagger} ( \bm{S} \cdot \bm{\sigma} ) c ,
	\label{eq:Hamiltonian_sd}
\end{align}
where $c = {}^t (c_{\uparrow}, c_{\downarrow})$ and $c^{\dagger} = (c^{\dagger}_{\uparrow}, c^{\dagger}_{\downarrow})$ 
are annihilation and creation operators of the electrons,
$a_{\bm{q}}$ and $a^{\dagger}_{\bm{q}}$ are those of magnons,
$m$ and $\mu$ are the mass and the chemical potential of the electrons,
$\omega_{\bm{q}} = J q^{2} + \Delta$ is the magnon dispersion 
with exchange stiffness $J$ and energy gap $\Delta$,
$\bm{S} = S \bm{n}\ (|\bm{n}| = 1)$ is the localized spin with magnitude $S$,
$\bm{\sigma} = (\sigma^{x},\sigma^{y},\sigma^{z})$ are Pauli matrices,
and $J_{sd}$ is the \textit{s-d} exchange coupling constant. 
 We consider low enough temperature and assume $S$ is constant. 
Hereafter we use $M \equiv J_{sd} S$ and $\bm{n}$ instead of $\bm{S}$. 
 For $V_{\rm imp}$, we consider both nonmagnetic and magnetic impurities, 
\begin{align}
	V_{\rm imp} (\bm{r})
	= u_{\rm i} \sum_{i} \delta (\bm{r} - \bm{R}_{i})
	+ u_{\rm s} \sum_{j} \bm{S}_{j} \cdot \bm{\sigma} \, \delta (\bm{r} - \bm{R}'_{j}),
	\label{eq:def_Vimp}
\end{align}
where $\bm{S}_{j}$ is the impurity spin located at position $\bm{R}'_{j}$.
 We average over the impurity positions, ${\bm R}_i$ and ${\bm R}_j'$, as usual, 
and the impurity spin directions, 
\begin{align}
	\overline{S^{\alpha}_{i} S^{\beta}_{j}}
	=
	\delta_{ij} \delta_{\alpha \beta} \times \left\{ 
	\begin{array}{cl} 
		\overline{S^{2}_{\perp}} & (\alpha = \beta = x,y) \\
		\overline{S^{2}_{z}} & (\alpha = \beta = z)
	\end{array}  \right.  . 
	\label{eq:average_impurityspins}
\end{align}
 The \textit{s-d} exchange interaction describes the exchange-splitting in the electron 
spectrum, and the electron-magnon scattering, 
\begin{align}
	\mathcal{H}_{sd}
	=&
	- M \int {\rm d} \bm{r} c^{\dagger} \sigma^{z} c
	+ \mathcal{H}_{\rm el-mag},
	\label{eq:decomp_Hsd}
	\\
	\mathcal{H}_{\rm el-mag}
	=&
	M \int {\rm d} \bm{r}
	\left[ 
		\frac{1}{s_{0}} a^{\dagger} a \, \hat{\sigma}^{z}
		-
		\sqrt{\frac{2}{s_{0}}}
		\left( a \hat{\sigma}^{-} + a^{\dagger} \hat{\sigma}^{+} \right)
	\right],
	\label{eq:def_Helmag}
\end{align}
where $s_{0} = S/ r_{0}^{3}$ is the spin density of the magnetization, 
$r_{0}$ the lattice constant, 
$\hat{\bm{\sigma}} = c^{\dagger} \bm{\sigma} c$,  
and $\hat\sigma^{\pm} = (\hat\sigma^{x} \pm i \hat\sigma^{y})/2$.
The total Hamiltonian is given by 
\begin{align}
	\mathcal{H}
	&= 
	\mathcal{H}_{{\rm el}} + \mathcal{H}_{{\rm mag}} + \mathcal{H}_{\rm el-mag},
	\label{eq:decomp_Hamiltonian}
	\\
	\mathcal{H}_{{\rm el}}
	&= 
	\int {\rm d} \bm{r}
	\left[ 
		\frac{1}{2m} (\partial_{i} c^{\dagger}) (\partial_{i} c)
		+ c^{\dagger} (V_{\rm imp} - \mu) c - M c^{\dagger} \sigma^{z} c \, 
	\right].
	\label{eq:def_H_el}
\end{align}

\subsection{Green's function}

The Green's functions of electrons $G_{\bm{k}\sigma}(i \varepsilon_{n})$ 
and magnons $D_{\bm{q}} (i \nu_{l})$ are given by 
\begin{align}
 G_{\bm{k}\sigma} (i\varepsilon_{n})
&= \frac{1}{i \varepsilon_{n} + \mu  - {\bm k}^2/2m + \sigma M - \Sigma_{\sigma}(i \varepsilon_{n})},
\label{eq:def_Green_el}
\\
 D_{\bm{q}} (i \nu_{l}) &= \frac{1}{i \nu_{l} - \omega_{\bm{q}} - \Pi_{\bm{q}}(i \nu_{l})},
\label{eq:def_Green_mag}
\end{align}
with Matsubara frequencies, $\varepsilon_n = (2n+1) \pi T$ and $\nu_l = 2\pi l T$, 
and self-energies, $\Sigma_{\sigma}(i \varepsilon_{n})$ and $\Pi_{\bm{q}}(i \nu_{l})$, 
for the electrons and magnons, respectively.

\begin{figure}[b]
	\centering
		\includegraphics[width=70mm]{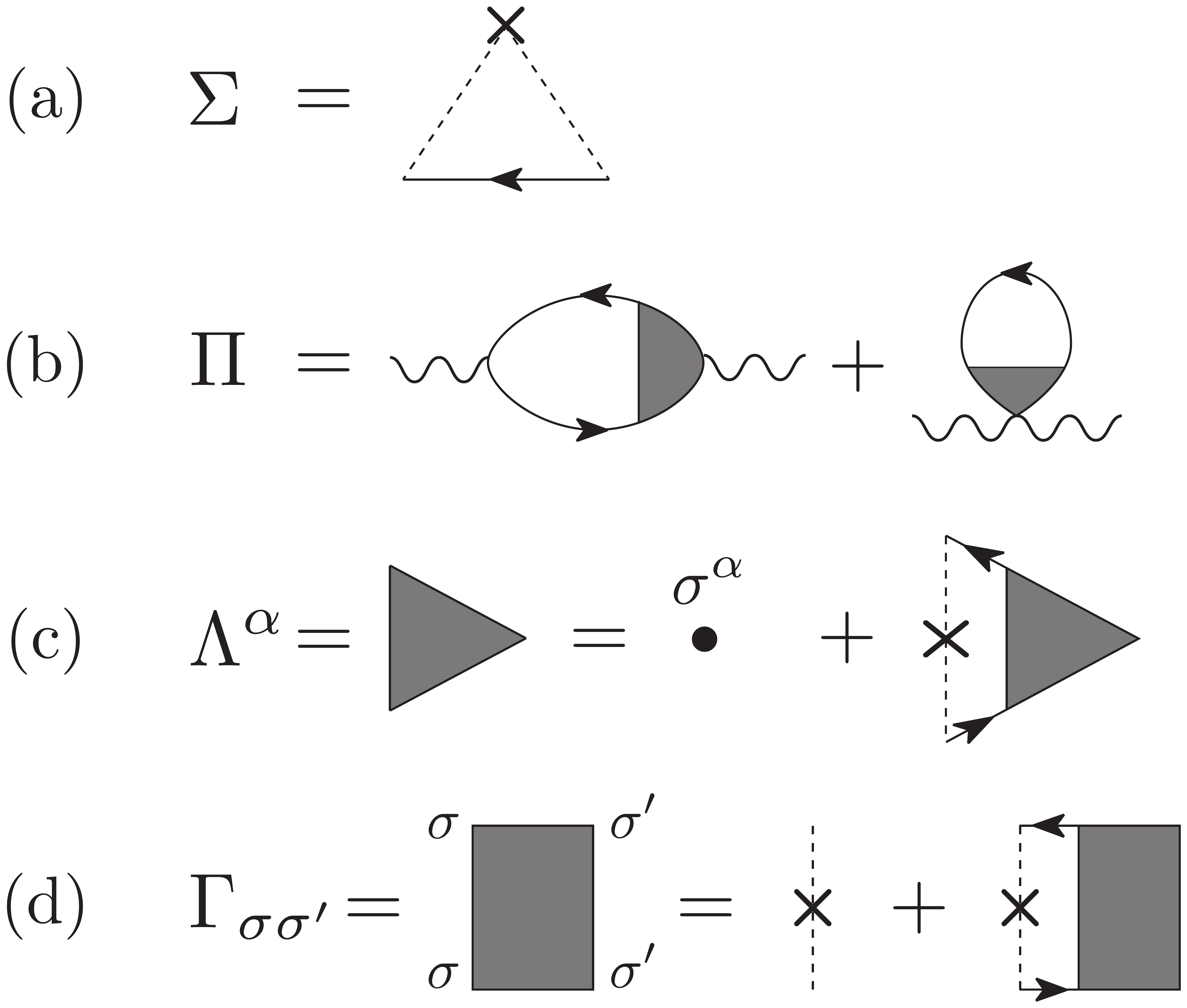}
	\caption{
		(a) Self-energy of electrons, $\Sigma$. (b) Self-energy of magnons, $\Pi$. 
	    (c) Spin vertex $\Lambda^\alpha$ renormalized by impurity-ladder corrections. 
		(d) Four-point vertex $\Gamma_{\sigma\sigma'}$, which we call the diffusion-type 
         vertex correction, or simply, the diffusion propagator. 
		 The solid (wavy) lines represent electron (magnon) propagators, 
		 and the dashed line with a cross represents impurity scattering.  
	}
	\label{fig:SE}
\end{figure}

We assume the electron self-energy is dominated by impurity scattering 
and treat it in the Born approximation [Fig.~1 (a)].
 Thus, 
$\Sigma_{\sigma}^{\rm R}(\varepsilon) = \Sigma_{\sigma}(\varepsilon + i0) = - i \gamma_{\sigma}$,
with 
\begin{align}
	\gamma_{\sigma}
	=
	\pi (\Gamma_{1} \nu_{\sigma} + \Gamma_{2} \nu_{\bar{\sigma}})
	\equiv
	\frac{1}{2 \tau_{\sigma}},
	\label{eq:def_gamma}
\end{align}
and 
\begin{align}
	\Gamma_{1} &= n_{\rm i} u_{\rm i}^{2} + n_{\rm s} u_{\rm s}^{2} \overline{S^{2}_{z}} , 
\ \ \ \ \  
    \Gamma_{2} = 2 n_{\rm s} u_{\rm s}^{2} \overline{S^{2}_{\perp}} . 
\label{eq:Gamma}
\end{align}
 Here, 
$n_{\rm i}$ ($n_{\rm s}$) is the concentration of nonmagnetic (magnetic) impurities, 
and $\nu_{\sigma}$ is the density of states of spin-$\sigma$ electrons.

The magnon self-energy comes from the electron-magnon scattering [Fig.~1 (b)].
Expanding with respect to the wave vector $\bm{q}$ and the frequency $\nu$ of magnons,
we write  
\begin{align}
	\Pi_{\bm{q}}(\nu + i0)
	&= 
	-\left[ \frac{\delta S}{S} + i \frac{\alpha}{z} \right] \nu
	- \delta J q^{2}
	+ \mathcal{O}(\nu^{2},q^{4}) . 
	\label{eq:def_selfenergy_mag}
\end{align}
 Here, $\delta S$, $\delta J$ and $z \equiv S/(S + \delta S)$ are the renormalization constants 
for spin, the exchange stiffness, and wave function, respectively, of the localized spins. 
 Also, $\alpha$ is the Gilbert damping constant calculated as \cite{Kohno2006}
\begin{align}
	\alpha
	=
	\pi n_{\rm s} u_{\rm s}^{2}
	\left[ 
		2 \overline{S^{2}_{z}} \nu_{\uparrow} \nu_{\downarrow}
		+
		\overline{S^{2}_{\perp}} (\nu_{\uparrow}^{2} + \nu_{\downarrow}^{2})
	\right]
	z / s_{0}.
	\label{eq:def_alpha}
\end{align}
 Here and hereafter, we assume the $s$-$d$ exchange coupling $M$ is much larger 
than the spin-relaxation rate \cite{Yamaguchi2017}. 

 As seen from ${\cal H}_{\rm el-mag}$ [Eq.~(\ref{eq:def_Helmag})], 
the natural expansion parameter in the electron-magnon problem is $s_0^{-1}$ (or $S^{-1}$). 
 In this paper, we focus on the leading contributions, which are ${\cal O}(s_{0}^{-1})$. 
(As seen below, we need two electron-magnon scattering vertices in the magnon-drag process, 
giving $\sim (s_{0}^{-1/2})^{2} = s_{0}^{-1}$.)
 Since $\delta S$ and $\delta J$ are ${\cal O}(s_{0}^{-1})$, 
we set $z=1$ and $\delta J = 0$ in the magnon Green's function.

\begin{figure}[t]
	\begin{center}
		\includegraphics[width=8.5cm]{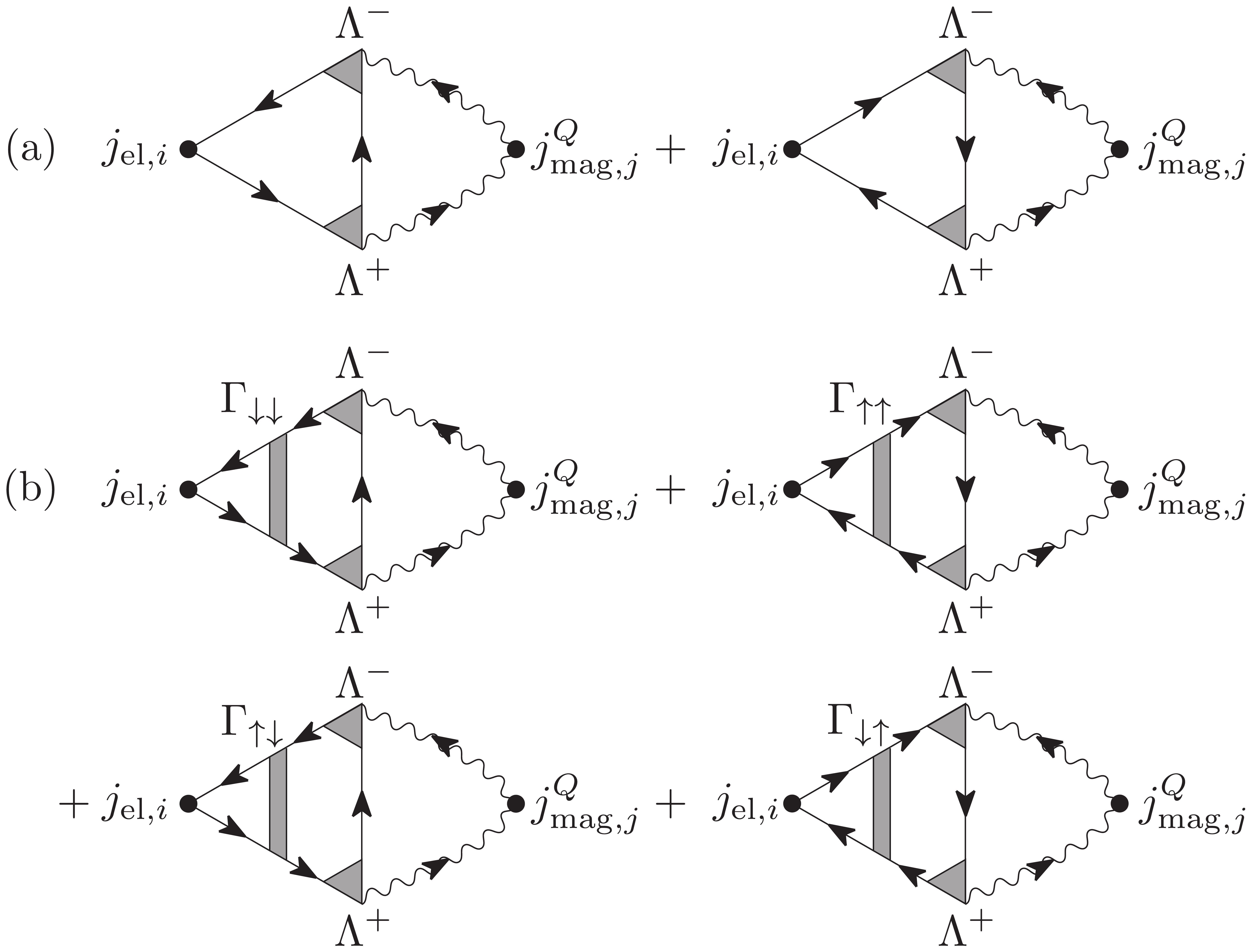}
	\end{center}
	\caption{
		Feynman diagrams for $K_{ij}(i \omega_{\lambda})$ [Eq.~(\ref{eq:def_Kij})], 
        which describe the magnon-drag processes.
		The solid (wavy) lines represent the electron (magnon) Green's functions. 
		(a) Processes for ${\bm Q}={\bm 0}$. 
		 The gray triangles are defined in Fig.~\ref{fig:SE} (c).         
        (b) Additional processes that contribute when ${\bm Q} \ne {\bm 0}$.  
         The gray square represents the diffusion propagator 
        $\Gamma_{\sigma\sigma'}$ defined in Fig.~\ref{fig:SE} (d). 
        The diagrams in (b) vanish for ${\bm Q}={\bm 0}$, but contribute for finite ${\bm Q}$ 
        and lead to Eq.~(\ref{discuss_jel_diffusive}). 
	}
	\label{fig:Kij}
\end{figure}

\section{Microscopic Calculation}\label{sec:calculation}

\subsection{Thermal linear-response theory}

To treat the temperature gradient in the linear response theory,
we introduce Luttinger's (fictitious) gravitational potential $\psi$,
which couples to the energy density $h({\bm r})$ of the system \cite{Luttinger1964}. 
 The coupling is described by the Hamiltonian, 
\begin{align}
 \mathcal{H}' 
 = \int {\rm d} \bm{r} \, h (\bm{r}) \psi (\bm{r}, t) . 
\label{eq:def_Hex}
\end{align}
 We consider the case, 
$\psi ({\bm r}, t) = \psi_{{\bm Q}, \omega} e^{i({\bm Q} \cdot {\bm r} - \omega t)}$, 
where ${\bm Q}$ and $\omega$ are the wave vector and the frequency of $\psi$, 
and write the linear response of a physical quantity $A$ to $\psi$ as 
\begin{align}
	\left< A \, \right>^{\psi}_{\omega}
	=
	- \left< A \, ; h(-{\bm Q}) \, \right>_{\omega+i0} \psi_{{\bm Q}, \omega} , 
	\label{eq:def_linear_responce}
\end{align}
where $h (-{\bm Q})$ is the Fourier component of $h({\bm r})$. 
The response function $\left< A \, ; B \right>_{\omega+i0}$ is obtained from 
\begin{align}
	\left< A \, ; B \, \right>_{i \omega_{\lambda}}
	\equiv 
	\int_{0}^{T^{-1}} {\rm d} \tau \, {\rm e}^{i \omega_{\lambda} \tau}
	\left< {\rm T}_\tau A (\tau) B \, \right>_{\rm eq},
	\label{eq:def_correlation_function}
\end{align}
by the analytic continuation, $i \omega_{\lambda} \to \omega + i0$, 
where $A$ and $B$ are arbitrary operators. 
 Here, $T$ is the temperature and
$\left< \cdots \right>_{\rm eq}$ represents the average in thermal equilibrium.
Hereafter we use $\left< \cdots \right>$ instead of $\left< \cdots \right>_{\rm eq}$ for simplicity.
 Using the continuity equation, 
\begin{align}
	\partial_{t} h(\bm{r}) + \partial_{i} j^Q_i = 0 , 
	\label{eq:def_continuity_equation_energy}
\end{align}
which defines the heat-current density $j^Q_i$,
we rewrite Eq.~(\ref{eq:def_linear_responce}) as a 
linear response to $(- \partial_{i} \psi)$ \cite{Kohno2016}, 
\begin{align}
 &  \left< A \, \right>^{\psi}_{\omega}
	=
	\frac{K_{i}(\omega + i0) - K_{i}(0)}{i \omega} 
    \left( - \partial_{j} \psi - \frac{\partial_{j} T}{T} \right)_{\bm Q} ,
	\label{eq:rewrite_linear_response}
	\\
 & \ \ K_{i} (i \omega_{\lambda})
	=
	\langle \, A \, ; \,  j^{Q}_{i} (-{\bm Q}) \rangle_{i \omega_{\lambda}} . 
	\label{eq:def_Ki}
\end{align}
 Here, we introduced the temperature gradient $\partial_i T$ through the combination, 
$- \partial_i \psi - \partial_i T/T$. 
 This is justified for operators $A$ of which the average vanishes naturally 
in the equilibrium state, 
where $\partial_i T / T + \partial_i \psi = 0$ holds \cite{Luttinger1964,Kohno2016}.   
 Therefore, the response to $(- \partial_i T /T)$ is obtained as 
the response to $(- \partial_i \psi)$ \cite{Luttinger1964}.

\subsection{Magnon-drag process}

 Specializing to the present model, Eq.~(\ref{eq:decomp_Hamiltonian}), we find from 
Eq.~(\ref{eq:def_continuity_equation_energy}) that 
the heat-current density  $j^{Q}_{i}$ consists of two parts, 
$j^{Q}_{i} = j^{Q}_{{\rm el},i} + j^{Q}_{{\rm mag}}$, one for the electrons 
($j^{Q}_{{\rm el},i}$) and one for magnons, 
\begin{align}
	j^{Q}_{{\rm mag},i}
= - J \left[ \dot{a}^{\dagger} (\partial_{i} a) + (\partial_{i} a^{\dagger}) \dot{a}  \right] , 
\label{eq:def_magnon_heatcurrent}
\end{align}
where $\dot{a} = \partial_{t} a$.

 In this paper, we are interested in the magnon-drag process,  
which corresponds to taking the magnon heat-current density 
$j^{Q}_{{\rm mag},i}$ for $j^{Q}_{i}$ in Eq.~(\ref{eq:def_Ki}). 
 As for $A$ in Eq.~(\ref{eq:def_Ki}), we focus on the electron (number) current density, 
\begin{align}
  j_{{\rm el}, i} &= \frac{\hbar}{2mi} \left[ \, c^\dagger (\partial_i c) - (\partial_i c^\dagger) c \, \right] . 
\end{align}
 Therefore, we consider 
\begin{align}
	& \left< \,  j_{{\rm el},i} \right>_{\rm drag}
	=
	\frac{K_{ij}(\omega + i0) - K_{ij}(0)}{i \omega} 
	\left( - \partial_{j} \psi - \frac{\partial_{j} T}{T} \right)  ,
	\label{eq:def_jel_psi}
\\ 
 & \ \ \  K_{ij}(i \omega_{\lambda})
= \langle \, j_{{\rm el},i} ({\bm Q}) \, ; \, j^{Q}_{{\rm mag},j} (-{\bm Q}) \rangle_{i\omega_\lambda} , 
 \label{eq:def_Kij}
\end{align}
i.e., the correlation function between the electron (number) current 
and the magnon heat current. 
 Here $j_{{\rm el},i} ({\bm Q})$ and $j^{Q}_{{\rm mag},j} (-{\bm Q})$ represent their respective 
Fourier components. 
 The combination $- \partial_j \psi - \partial_j T/T $ in Eq.~(\ref{eq:def_jel_psi}) 
indicates that the current vanishes in the equilibrium state, 
in which $- \partial_j \psi - \partial_j T/T = 0$ (Einstein-Luttinger relation) holds. 
 We will verify this form by an explicit calculation in Secs.~V-A and V-B.

 The relevant magnon-drag processes are shown diagrammatically in Fig.~\ref{fig:Kij} (a). 
 These are the leading contribution with respect to $1/s_0$, and expressed as 
\begin{widetext}
\begin{align}
	K_{ij}(i \omega_\lambda)
&= \frac{2M^2}{s_0} \, T \sum_{l, {\bm q}} u_j \left\{ 
	\left( i \nu_l + \frac{i \omega_\lambda}{2} \right)  
	D_{\bm q} (i \nu_l + i \omega_\lambda)  D_{\bm q} (i \nu_l) 
    - \frac{1}{2} \left[ D_{\bm q} (i \nu_l + i \omega_\lambda) + D_{\bm q} (i \nu_l) \right] \right\} 
    \mathcal{E}_i  , 
	\label{eq:Kij_Green}
\end{align}
where $u_{i} = 2 J q_{i}$ is the magnon velocity, 
$\omega_{\lambda}$ is the Matsubara frequency of the external perturbation $\psi$, 
and we have set $\bm{Q} = {\bm 0}$ for simplicity. 
 The terms linear in $D_{\bm q}$ are \lq\lq corrections'' 
arising from the $\delta$-function in the relation \cite{Kohno2016}, 
\begin{eqnarray}
  \langle \, {\rm T}_\tau \, a \, (\tau) \, \dot a^\dagger \, \rangle 
= -\langle \, {\rm T}_\tau \, \dot a \, (\tau) \, a^\dagger \, \rangle 
= \frac{d}{d\tau} D(\tau) + \delta (\tau) . 
\end{eqnarray}
 These terms, combined with the first term ($\sim D_{\bm q} D_{\bm q}$) in the curly brackets, lead to 
$\{ \cdots \} = \{ \omega_{\bm q} 
   + \frac{1}{2} [ \Pi_{\bm q} (i \nu_l + i \omega_\lambda) + \Pi_{\bm q} (i \nu_l) ] \} 
  D_{\bm q} D_{\bm q}$. 
 This amounts to making a replacement, $i\nu_l + i\omega_\lambda /2 \to \omega_{\bm q}$, 
in the first term if the self-energies are neglected.

 The last factor $\mathcal{E}_{i}$ in Eq.~(\ref{eq:Kij_Green}) is the electron part 
coming from the electron triangles in Fig.~\ref{fig:Kij} (a), 
\begin{align}
	\mathcal{E}_{i}
	&= 
	T \sum_{n,\bm{k}}
	v_{i}
	\bigl[ 
		G_{\bm{k} \downarrow} (i \varepsilon_{n}+ i \omega_{\lambda}) \, 
		\Lambda^{-}_{\downarrow \uparrow} \, 
		G_{\bm{k}-\bm{q}, \uparrow} (i \varepsilon_{n} - i \nu_{l}) \, 
		\Lambda^{+}_{\uparrow \downarrow} \, 
		G_{\bm{k} \downarrow} (i \varepsilon_{n})
\nonumber \\
 & \hskip 16mm 
    +  G_{\bm{k} \uparrow} (i \varepsilon_{n}+ i \omega_{\lambda}) \, 
		\Lambda^{+}_{\uparrow \downarrow} \, 
        G_{\bm{k}+\bm{q}, \downarrow} (i \varepsilon_{n}+ i \nu_{l} + i \omega_{\lambda}) \, 
		\Lambda^{-}_{\downarrow \uparrow} \, 
		G_{\bm{k} \uparrow} (i \varepsilon_{n})
	\bigr],
	\label{def_Ei}
\end{align}
\end{widetext}
where $v_{i} = k_{i}/m$ is the electron velocity and 
$\Lambda_{\sigma \sigma'}^\pm$'s are the renormalized spin ($\sigma^\pm$) vertices; 
see Appendix \ref{sec:vertexcorrection} for the definition.
 After the analytic continuations, $i\nu_l \to \nu$ and  $i \omega_{\lambda} \to \omega + i0$,
an expansion is made with respect to $\omega$ and/or $\nu$. 
 From Eq.~(\ref{eq:def_jel_psi}), we are primarily interested in the $\omega$-linear terms. 
 The factor $\omega$ comes either from the magnon part or from the electron part. 
 Hence we write  
\begin{widetext}
\begin{align}
	& K_{ij} (\omega + i0) - K_{ij} (0)
	\notag \\
	&\simeq 
	\frac{2M^{2}}{s_{0}} \frac{i \omega}{2\pi} 
    \Biggl\{ 
	\int {\rm d} \nu
	\left( - \frac{\partial n}{\partial \nu} \right) \nu 
	\sum_{\bm{q}} u_{j}  
    D_{\bm{q}}^{\rm R} (\nu) D_{\bm{q}}^{\rm A} (\nu) \, 
    \mathcal{E}^{(2)}_{i}
 - 2 \int {\rm d} \nu \, n(\nu) \nu 
	\sum_{\bm{q}} u_{j}  
	\Im \left[ D_{\bm{q}}^{\rm R} (\nu) D_{\bm{q}}^{\rm R} (\nu) \, 
    \frac{\mathcal{E}^{(1)}_{i}}{-i\omega} \right] 
\nonumber \\
& \hskip 18mm 
 - \frac{1}{2}\int {\rm d} \nu
	\left( - \frac{\partial n}{\partial \nu} \right)  
	\sum_{\bm{q}} u_{j}  
    \left[ D_{\bm{q}}^{\rm R} (\nu) + D_{\bm{q}}^{\rm A} (\nu) \, \right] 
    \mathcal{E}^{(2)}_{i}
 + 2 \int {\rm d} \nu \, n(\nu)  
	\sum_{\bm{q}} u_{j}  
	\Im \left[ D_{\bm{q}}^{\rm R} (\nu) \, 
    \frac{\mathcal{E}^{(1)}_{i}}{-i\omega} \right] 
    \Biggr\} , 
	\label{eq:anacon_Kij}
\end{align}
\end{widetext}
where $n(\nu) = ( e^{\nu / k_{\rm B} T} - 1)^{-1}$ is the Bose-Einstein distribution function. 
 The terms in the second line are the corrections mentioned above. 
$\mathcal{E}^{(1)}_{i}$ is obtained from $\mathcal{E}_{i}$ by the analytic continuation, 
$i(\nu_l + \omega_\lambda) \to \nu + \omega + i0$ and $i\nu_l \to \nu + i0$, 
and $\mathcal{E}^{(2)}_{i}$ by 
$i(\nu_l + \omega_\lambda) \to \nu + \omega + i0$ and $i\nu_l \to \nu - i0$. 
 In the term with $\mathcal{E}^{(2)}_{i}$, $\omega$ is picked up from the magnon part, 
whereas in the term with $\mathcal{E}^{(1)}_{i}$, $\omega$ is obtained from the electron part.

 At this point, it is worth noting that not only $D^{\rm R} D^{\rm A}$ 
but also $D^{\rm R} D^{\rm R}$ appears in Eq.~(\ref{eq:anacon_Kij}) 
for the pair of magnon propagators. 
 This is not surprising in diagrammatic calculations as being done here, 
but seems incompatible with the spin-motive force picture, 
in which there should be a causal relationship between the magnetization 
dynamics and the resulting current (see Sec.~IV).

 We retain low-order terms with respect to $\nu$, 
which is justified because the magnon energy $\nu$ is typically small 
compared to the electron Fermi energy. 
 Deferring the details to Appendix B, the electron part has been calculated as
\begin{align}
	\mathcal{E}^{(2)}_{i}
	&=
	 \frac{1}{(2M)^2} 
	\frac{ \sigma_{\uparrow} - \sigma_{\downarrow} }{e^2} 
      \{ 2\beta_{\rm el} \nu - i \omega (1 + i \beta_{\rm el}) \} q_i , 
	\label{result_Ei_2}
\\
	\mathcal{E}^{(1)}_{i}
	&\simeq  
	\frac{1}{(2M)^{2}}  
	\frac{ \sigma_{\uparrow} - \sigma_{\downarrow}}{e^2} (-i\omega q_i) , 
	\label{result_Ei_1}
\end{align}
where $\sigma_{\sigma} = e^{2} (v_{ {\rm F}\sigma}^{2}/3) \nu_\sigma \tau_{\sigma}$
is the conductivity of electrons with spin $\sigma$ ,  
\begin{align}
	\beta_{\rm el}
	=
	\frac{\pi n_{\rm s} u_{\rm s}^{2}}{M}
	\left[ 
		( \overline{S^{2}_{\perp}} + \overline{S^{2}_{z}} ) \, \nu_{+}
	+ P^{-1} ( \overline{S^{2}_{\perp}} - \overline{S^{2}_{z}} ) \, \nu_{-}
	\right],
	\label{def_beta}
\end{align}
is the so-called $\beta$ parameter that parametrizes the dissipative corrections  
to the spin-transfer torque \cite{Kohno2006,Duine2007} 
and to the Berry-phase spin-motive force \cite{Duine2008,Tserkovnyak2008}. 
 We define $\nu_{\pm} = \nu_{\uparrow} \pm \nu_{\downarrow}$ and
$P = (\sigma_{\uparrow} - \sigma_{\downarrow}) / (\sigma_{\uparrow} + \sigma_{\downarrow})$. 
 For the present purpose, we can discard the $\omega$-linear term in $\mathcal{E}^{(2)}_{i}$. 
 It will be used in Sec.~IV when we discuss the spin-motive force.

 The magnon part is calculated by using 
\begin{align}
   \frac{1}{2\pi}
	\int {\rm d} \nu
	\left( - \frac{\partial n}{\partial \nu} \right)
	\nu^{2}
	\sum_{\bm{q}}
	u_{i} q_{j} D^{\rm R}_{\bm{q}} D^{\rm A}_{\bm{q}}
	& =
	\frac{1}{2\alpha} T \mathcal{S}_{\rm mag} \delta_{ij},
	\label{int_magnon_RA}
	\\
	\frac{1}{\pi}
	\int {\rm d} \nu
	n(\nu) \, \nu
	\sum_{\bm{q}}
	u_{i} q_{j}
	\Im \left[ \left( D^{\rm R}_{\bm{q}} \right)^{2} \right]
	& =
	\mathcal{E}_{\rm mag} \delta_{ij} ,
	\label{int_magnon_RR}
	\\
	\frac{1}{\pi}
	\int {\rm d} \nu
	n(\nu) \, 
	\sum_{\bm{q}} u_{i} q_{j}
	\Im \left[ D^{\rm R}_{\bm{q}} \right]
	& =
	\Omega_{\rm mag} \delta_{ij} ,
	\label{int_magnon_RR_corr}
\end{align}
where $\mathcal{E}_{{\rm mag}} = \sum_{\bm{q}} \omega_{\bm{q}} n(\omega_{\bm{q}})$ is the energy density,  
$\Omega_{\rm mag} = k_{\rm B} T \sum_{\bm q} \ln ( 1 - e^{- \hbar \omega_{\bm q}/k_{\rm B}T} )$ 
is the thermodynamic potential density, and 
$\mathcal{S}_{\rm mag} = - \partial \, \Omega_{\rm mag} / \partial T$ 
is the entropy density of magnons.
 Thus the magnon-drag electron (number) current is obtained as 
\begin{widetext}
\begin{align}
	\left< j_{{\rm el},i} \right>_{\rm drag}
&= - \frac{1}{2 s_{0}} \frac{\sigma_{\uparrow} - \sigma_{\downarrow}}{e^{2}}
	\left[  \mathcal{E}_{{\rm mag}} - \frac{\beta_{{\rm el}}}{\alpha} T \mathcal{S}_{{\rm mag}} - \Omega_{\rm mag}  \right]
 \left( - \partial_{i} \psi \right) 
\nonumber \\
&= - \frac{1}{2 s_{0}} \frac{\sigma_{\uparrow} - \sigma_{\downarrow}}{e^{2}}
	\left(  1 - \frac{\beta_{{\rm el}}}{\alpha} \right) T \mathcal{S}_{{\rm mag}} 
 \left( - \partial_{i} \psi \right) , 
	\label{result_jel_micro}
\end{align}
\end{widetext}
where we used $\Omega_{\rm mag} = {\cal E}_{\rm mag} - T {\cal S}_{\rm mag}$. 
 Note that $\Omega_{\rm mag}$, which arises as \lq\lq corrections'' here, turned the energy 
${\cal E}_{\rm mag}$ into the entropy $T {\cal S}_{\rm mag}$, and the result depends 
on magnons only through their entropy. 
 This is the main result of this paper.

\subsection{Result}\label{sec:result}

 A physical result is obtained by replacing $\partial_i \psi$ by $\partial_i T/T$,
\begin{align}
	\left< {\bm j}_{\rm el} \right>_{\rm drag}
&= - \frac{1}{2 s_{0}}
	\frac{\sigma_{\uparrow} - \sigma_{\downarrow}}{e^2}
	 \left( 1	- \frac{\beta_{{\rm el}}}{\alpha} \right) T \mathcal{S}_{{\rm mag}} 
	\left( - \frac{ {\bm \nabla} T}{T} \right)  
	\label{discuss_result_jel_micro}
\\ 
&= - \frac{1}{2 s_{0}} \frac{\sigma_{\uparrow} - \sigma_{\downarrow}}{e^{2}}
	\left(  1 - \frac{\beta_{{\rm el}}}{\alpha} \right) {\bm \nabla} \Omega_{{\rm mag}} . 
	\label{result_jel_micro_grad}
\end{align}
 In the second line, we noted 
${\cal S}_{\rm mag} = - \partial \Omega_{\rm mag} /\partial T$ 
and assumed that $\Omega_{\rm mag}$ is position (${\bm r}$) dependent only 
through the local temperature, $T=T({\bm r})$.

  The obtained magnon-drag current, Eq.~(\ref{result_jel_micro_grad}), is 
proportional to $\sigma_{\uparrow} - \sigma_{\downarrow}$ \cite{com1}. 
 This indicates that the magnons exert on the electrons a spin-dependent force, 
\begin{align}
 {\bm F}_\sigma 
&= - \frac{\sigma}{2 s_{0}} 
 \left(  1 - \frac{\beta_{\rm el}}{\alpha} \right) {\bm \nabla} \Omega_{\rm mag} , 
\label{eq:Fs} 
\end{align}
where $\sigma = 1$ or $-1$ depending on the electron spin projection, 
$\sigma = \, \uparrow$ or $\downarrow$. 
 Some discussion will be given in Sec.~IV in relation to the spin-motive force.

 Equation (\ref{eq:Fs}) has the form of total gradient, suggesting 
that it is of diffusive nature and is induced 
by a spin-dependent, nonequilibrium chemical potential,  
\begin{align}
 \delta \mu_\sigma 
&= \frac{\sigma}{2 s_{0}} 
	\left(  1 - \frac{\beta_{\rm el}}{\alpha} \right) \delta \Omega_{{\rm mag}} , 
\label{eq:delta_mu_sigma}
\end{align}
where $\delta \Omega_{{\rm mag}}$ is the deviation of $\Omega_{{\rm mag}}$ 
from its thermal-equilibrium value. 
 In Sec.~V-B, we will give a further analysis that supports this picture 
of the spin chemical potential.

\section{Phenomenology based on spin-motive force}\label{sec:phenomenology}

 In this section, we revisit the phenomenology based on the spin-motive force 
along the lines of Refs.~\cite{Lucassen2011,Flebus2016}, 
and compare the result with the microscopic result. 
 The physical pictures that emerge from the microscopic study are also discussed.

 When the magnetization vector ${\bm n}$ varies in space and time, 
it exerts a spin-dependent force, $\pm F_i$, on electrons, where 
\begin{align}
 F_{i} =
 \frac{\hbar}{2}
 \left[ \bm{n} \cdot \left( \dot {\bm n} \times \partial_{i} \bm{n} \right)
  - \beta \, \dot {\bm n} \cdot \partial_{i} \bm{n}  \right] . 
	\label{def_Fi}
\end{align}
 This is called the spin-motive force. 
 The first term is the \lq\lq Berry phase term'' and the second term with a 
dimensionless coefficient $\beta$ is the dissipative correction, 
which we call the $\beta$-term \cite{Duine2008, Tserkovnyak2008, Shibata2011}. 
 ($\beta$ is equal to $\beta_{\rm el}$ [Eq.~(\ref{def_beta})], but we continue to use these notations; 
$\beta_{\rm el}$ for the microscopically-calculated one,              
and $\beta$ for the phenomenologically-introduced one.)  
 These effects are reciprocal to the current-induced spin torques; 
the former is reciprocal to the spin-transfer torque, 
and the latter to its dissipative correction.

 Spin waves, or magnons, can also be the origin of the spin-motive force. 
 Although they are fluctuations, they will induce a net electron current 
\begin{align}
 \left< j_{{\rm el}, i} \right>_{\rm smf}
= \frac{\sigma_{\uparrow} - \sigma_{\downarrow}}{e^2} \langle F_i \rangle , 
	\label{def_jismf}
\end{align}
if the average survives, $\left< F_{i} \right> \neq 0$. 
 This will contribute to the magnon-drag electron current. 
 Here we assume a uniformly magnetized state at equilibrium, $\bm{n}|_{\rm eq} = \hat{z}$, 
and consider small fluctuations $\delta \bm{n}$ around it, 
such that $\bm{n} = \hat{z} + \delta \bm{n}$. 
 With magnon operators, $\{ a , a^\dagger \} = (s_{0}/2)^{1/2}(\delta n_{x} \pm i \delta n_{y})$, 
we rewrite $F_{i}$ as
\begin{align}
 F_i &=  \frac{i}{2 s_0}
	\left[ - \dot a^\dagger \partial_i a + (\partial_i a^\dagger) \dot a \right] 
	- \frac{\beta}{2s_0}
	\left[ \dot a^\dagger \partial_i a + (\partial_i a^\dagger) \dot a \right] . 
\label{eq:Fi_SW}
\end{align}
 As noted previously \cite{Lucassen2011,Flebus2016}, 
the second term is essentially the magnon heat current  $j^{Q}_{{\rm mag},i}$  [Eq.~(\ref{eq:def_magnon_heatcurrent})]. 
 Here we note that the first term is expressed  by the magnon energy, 
$h_{\rm mag} = i( a^\dagger \dot a - \dot a^\dagger a)/2$, and the magnon current 
$j_{{\rm mag}, i} = -iJ  [ a^\dagger \partial_i a - (\partial_i a^\dagger) a ] $. 
 Thus, 
\begin{align}
 F_i  
&=  \frac{1}{2 s_0} \left\{ \partial_i  h_{\rm mag}    
    + \frac{1}{2J} \frac{\partial}{\partial t}  j_{{\rm mag}, i} 
	+ \frac{\beta}{J}  j^{Q}_{{\rm mag},i}  \right\} . 
\label{eq:Fi_phen}
\end{align}

 Let us evaluate each term in Eq.~(\ref{eq:Fi_phen}) for a steady state with 
a temperature gradient. 
 Since the first term has a spatial derivative $\partial_i$, we evaluate it in the local 
equilibrium state as $\langle h \rangle = {\cal E}_{\rm mag}(T)$, 
which depends on ${\bm r}$ through the local temperature $T = T({\bm r})$. 
 This leads to 
$\partial_i  \langle h_{\rm mag} \rangle = (\partial {\cal E}_{\rm mag}/\partial T)(\partial_i T)$. 
 The second term vanishes in the steady state because of the overall time derivative. 
 The third term is evaluated as 
$\langle j^{Q}_{{\rm mag},i} \rangle = - \kappa \partial_i T$ 
with the magnon heat conductivity $\kappa$. 
 This is calculated using, e.g., the Kubo-Luttinger formula as \cite{Imai2018}  
\begin{align}
 \kappa   
&= \frac{1}{T} \int \frac{d\nu}{2\pi}  \left( - \frac{\partial n}{\partial \nu} \right) \nu^2 
  \sum_{\bm q} u_x^2 D^{\rm R}_{\bm{q}} D^{\rm A}_{\bm{q}}
= \frac{J}{\alpha} \mathcal{S}_{\rm mag} ,
\label{eq:kappa}
\end{align}
where we used Eq.~(\ref{int_magnon_RA}). 
 This expression for $\kappa$ in terms of magnon entropy also follows from 
an intuitive argument. 
 Following Drude, one may express the magnon heat-current density at position $x$ 
as \cite{Aschcroft}  
\begin{align}
 j_x^Q (x) 
&= \frac{1}{2} \sum_{\bm q} u_x \omega_{\bm q} \left[ n (x-u_x \tau) - n (x+u_x \tau) \right] , 
\end{align}
where $n(x)$ is the Bose distribution function defined with a local temperature 
$T(x)$, $\tau = (2\alpha\omega_{\bm q})^{-1}$ is the lifetime of magnons, 
and the temperature gradient is assumed in the $x$ direction. 
 The first term represents the energy flow from the left region, and the second term  
from the right, which are due to magnons that experienced their last collision at 
$x \pm u_x \tau$; 
the factor 1/2 is there because half of magnons at $x \pm u_x \tau$ 
(namely, those with $q_x>0$ or $q_x<0$) propagate to $x$. 
 Expanding as $n (x-u_x \tau) - n (x+u_x \tau) 
\simeq - 2 u_x \tau (\partial n/\partial T) (\partial T/\partial x)$ 
and using Eq.~(\ref{eq:Omega_mag_identity_1}), one has 
\begin{align}
 \kappa 
&= \frac{1}{2\alpha} \frac{\partial}{\partial T} \sum_{\bm q} u_x^2 n (\omega_{\bm q}) 
= \frac{J}{\alpha} {\cal S}_{\rm mag} , 
\end{align}
in agreement with Eq.~(\ref{eq:kappa}).

 Taken together, we obtain  
\begin{align}
 \langle F_i \rangle 
&=  \frac{1}{2 s_0} \left\{ - \frac{\partial {\cal E}_{\rm mag}}{\partial T}  
	+ \frac{\beta}{\alpha}  {\cal S}_{\rm mag} \right\} (- \partial_i T)  . 
\label{eq:phen_result}
\end{align}
 The same result has been obtained by other methods; see Appendix \ref{sec:pheno_anal}. 
 Therefore, we may conclude that any (phenomenological) theories starting from the 
spin-motive force lead to Eq.~(\ref{eq:phen_result}). 
 The first term is somewhat different from the one obtained in Ref.~\cite{Flebus2016}, 
and gives a slight revision to it (see Appendix C-3).

 We now compare Eq.~(\ref{eq:phen_result}) with the microscopic result,  Eq.~(\ref{discuss_result_jel_micro}). 
 One readily sees a disagreement in the first term, 
namely, the entropy ${\cal S}_{\rm mag}$ appears in the microscopic result 
instead of $\partial {\cal E}_{\rm mag}/\partial T$ in the phenomenological result.

\begin{figure}[t]
	\begin{center}
	\includegraphics[width=8.5cm]{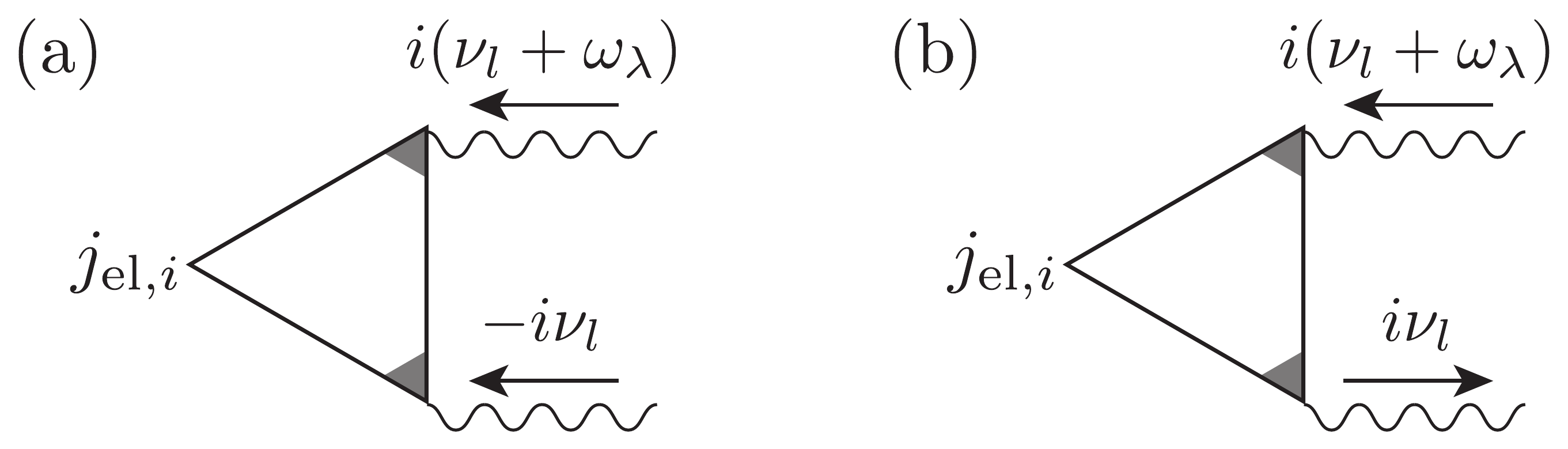}
	\end{center}
	\caption{ Feynman diagrams for the electric current induced by magnetization dynamics. 
     Arrows in the electron lines (solid lines) are suppressed for simplicity. 
	 (a) Nonlinear response to the (classical) magnetization dynamics. 
         The wavy lines represent the perturbations due to (classical) magnetization. 
         Because of causality (retarded response), the incoming Matsubara frequencies should satisfy 
         the conditions, $\nu_l + \omega_\lambda > 0$ and $- \nu_l > 0$ \cite{Kohno2007}. 
     (b) Part of the diagram of the present magnon-drag process (Fig.~2). 
          The wavy lines represent (quantum) magnon propagators. 
          Note that the flow of the Matsubara frequency in the lower magnon line is reversed 
          compared to (a). 
          The same causality relation as (a) leads to the analytic continuation, 
         $D(i\nu_l + i\omega_\lambda) D(i\nu_l) \to D^{\rm R} (\nu + \omega) D^{\rm A} (\nu)$, 
         for the pair of magnon propagators, 
         and this is associated with ${\cal E}^{(2)}_i$ given by Eq.~(\ref{result_Ei_2}). 
	}
	\label{fig:smf}
\end{figure}

 To identify the the origin of the difference, let us look at the Feynman diagram. 
 To calculate the spin-motive force, one calculates the electric current induced by magnetization dynamics \cite{Duine2008}. 
 This can be done by considering small fluctuations $\delta {\bm n}$ around the uniform magnetization, 
and look at the second-order (nonlinear) response to $\delta {\bm n}$ \cite{Kohno08}. 
 This is expressed diagrammatically in Fig.~\ref{fig:smf} (a), 
and the response function is given by ${\cal E}^{(2)}$ in Eq.~(\ref{result_Ei_2}). 
 Therefore, the induced current is calculated as  
\begin{align}
 \langle j_{{\rm el}, i} \rangle 
&= \frac{2M^2}{s_0} {\cal E}^{(2)}_i a_{{\bm q},\nu+\omega} a^*_{-{\bm q},-\nu} , 
\end{align}
where $\{ a, a^* \}$ is a classical (c-number) counterpart of $\{ a, a^\dagger \}$ 
defined just above Eq.~(\ref{eq:Fi_SW}), and the subscripts indicate their wave vector and frequency.  
 This leads to Eq.~(\ref{eq:Fi_SW}), hence to Eq.~(\ref{def_Fi}). 
 Therefore, the spin-motive force is described by the $\nu$- and $\omega$-linear terms in ${\cal E}^{(2)}_i$.  
 The appearance of ${\cal E}^{(2)}_i$ (originally from the magnon-drag calculation) 
in the nonlinear response here is due to the matching of the causality relationship; 
see Fig.~\ref{fig:smf} (b) and the caption thereof.

 On the other hand, in the present magnon-drag process, the first term comes from the 
$\omega$-linear term in ${\cal E}^{(1)}_i$, not from the $\omega$-linear term 
in ${\cal E}^{(2)}_i$;  the latter is irrelevant for the magnon-drag DC electron current. 
 Since ${\cal E}^{(1)}_i$ is accompanied by $D^{\rm R} D^{\rm R}$ (not $D^{\rm R} D^{\rm A}$), 
the physical interpretation of this term (in the magnon-drag current) does not necessarily 
rely on the causal relationship to the magnetization dynamics. 
 In fact, the spin-transfer process may be understood to occur 
in the quasi- or local-equilibrium situation, 
as will be discussed in the paragraph containing Eq.~(\ref{eq:reaction}).

\section{Spin chemical potential}\label{sec:discussion}

 In this section, we give an alternative argument that introduces a spin chemical potential. 
 This is intended to complement the heuristic discussion in Sec.~III-C.

 Our strategy here is as follows. 
 From the viewpoint of microscopic theory, statistical quantities such as 
the chemical potential and temperature, which characterize the distribution function, 
cannot be easily handled. 
 Instead, we can disturb the system by \lq\lq mechanical'' perturbations 
(which are described by the Hamiltonian and thus controllable theoretically) 
and then observe the result. 
 By examining how the distribution function is deformed, 
we may read off the change of statistical parameters such as chemical potential and temperature. 
 For example, an inhomogeneous potential (or electric field) induces a density modulation. 
 This effect is described by an inhomogeneous change of chemical potential, 
and appears in the current as a diffusion current \cite{Shibata2011}.

 In the following, we examine the possibility that the magnon-drag effects are described 
in a similar manner. 
 We first illustrate the procedure using a simple model (Sec.~V-A), 
and then consider the present problem of magnon-drag process (Sec.~V-B). 
 In both cases, we take the $\psi$ field as a mechanical perturbation.

\subsection{Electron-only process: Effective temperature}

\begin{figure}[t]
	\begin{center}
		\includegraphics[width=8.5cm]{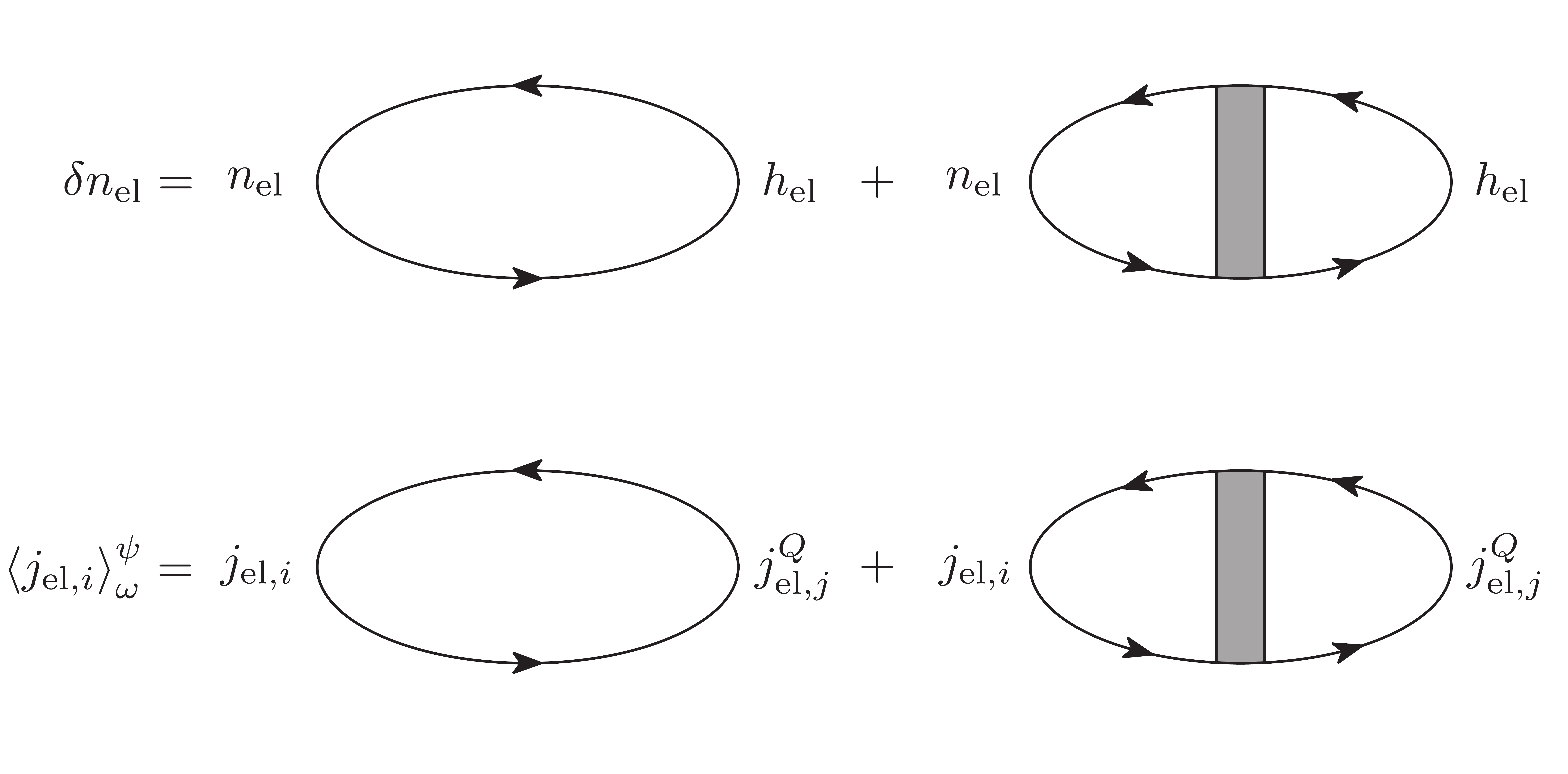}
	\end{center}
	\caption{
		Feynman diagrams for the electron density $\delta n_{\rm el}$ [Eq.~(\ref{nel_psi})] 
     and the current density $j_{{\rm el},i}$ [Eq.~(\ref{jel_psi})] induced by $\nabla T$.  
		$h_{{\rm el}}$ and $j_{{\rm el},i}^{Q}$ are the Hamiltonian density 
		and heat current density, respectively, of the conduction electrons.
		The shaded rectangle represents the diffusion-type ladder vertex correction 
        due to impurities [Fig.~1(d)], which describes diffusive motion of the electrons.
	}
	\label{fig:nel_jel}
\end{figure}

 We begin by reviewing the relation between the gravitational field $\partial_i \psi$
and temperature gradient, $\partial_i T$. 
 For simplicity, we consider a (spin-unpolarized) free electron system subject to nonmagnetic impurities, 
forgetting about magnons and even the magnetization (exchange splitting). 
 We calculate the electron density $\delta n_{{\rm el}}$ and current density 
$\left< j_{{\rm el},i} \right>^{\psi}_{\omega}$ induced by the disturbance $\psi$ having finite ${\bm Q}$ (and $\omega$). 
 In this case, it is essential to consider the diffusion-type vertex 
corrections [Fig.~1 (d)], hence the diagrams shown in Fig.~\ref{fig:nel_jel}.
 The results are given by 
\begin{align}
	\delta n_{{\rm el}}
	=&
	\int {\rm d} \varepsilon
	\left( - \frac{\partial f}{\partial \varepsilon} \right)
	\varepsilon \, \eta (\varepsilon),
	\label{nel_psi}
	\\
	\left< j_{{\rm el},i} \right>^{\psi}_{\omega}
	=&
	\int {\rm d} \varepsilon
	\left( - \frac{\partial f}{\partial \varepsilon} \right) \varepsilon
	\sigma(\varepsilon) (- \partial_{i} \psi)_{\bm{Q}}
	\notag \\
	& -
	\int {\rm d} \varepsilon
	\left( - \frac{\partial f}{\partial \varepsilon} \right) \varepsilon
	D (\varepsilon) \left[ \partial_{i} \eta (\varepsilon) \right]_{\bm{Q}},
	\label{jel_psi}
\end{align}
where 
\begin{align}
	\eta (\varepsilon)
	=&
	- \nu (\varepsilon)
	\frac{D(\varepsilon) Q^{2}}{D(\varepsilon) Q^{2} - i \omega} \psi_{\bm{Q}} , 
	\label{def_deltan}
\end{align}
describes \lq\lq diffusive corrections'' which arise since ${\bm Q}$ is finite. 
 We defined $e^2 \sigma(\varepsilon) = e^2 (v_{\rm F}^{2}/3) \nu(\varepsilon) \tau (\varepsilon) $
and $D(\varepsilon) = (v_{\rm F}^{2}/3) \tau(\varepsilon)$,  
which are the Boltzmann conductivity and the diffusion constant, respectively, 
evaluated at energy $\varepsilon$ (measured from the chemical potential $\mu$).

 If we consider a local modification of temperature, 
$T \to T + \delta T (\bm{r})$, the electron density changes by  
\begin{align}
	\delta n_{{\rm el}}
	&= 
	\int {\rm d} \varepsilon \, \nu(\varepsilon) 
	[ \, f (\varepsilon; T + \delta T) - f (\varepsilon; T) \, ] 
	\notag \\
	& \simeq 
	\frac{\delta T}{T}
	\int {\rm d} \varepsilon \, \nu(\varepsilon) \, \varepsilon  
	\left( - \frac{\partial f}{\partial \varepsilon} \right) . 
	\label{def_deltaT}
\end{align}
 In the \lq slow' limit $\omega \to 0$, Eq.~(\ref{nel_psi}) may be compared with Eq.~(\ref{def_deltaT}), and we may identify the effective temperature change 
$\delta T_{\bm{Q}}$ by 
\begin{align}
	- \lim_{\omega \to 0} \frac{D(\varepsilon) Q^{2}}{D(\varepsilon) Q^{2} - i \omega}
	\psi_{\bm{Q}}
	=
	\frac{\delta T_{\bm{Q}}}{T} . 
	\label{Einstein_relation}
\end{align}
 This is nothing but the Einstein-Luttinger relation, $\psi_{\bm Q} + \delta T_{\bm Q}/T = 0$, 
that holds in the equilibrium state (under a static potential, $\psi_{\bm Q}$). 
 Using this $\delta T_{\bm Q}$, we may rewrite Eq.~(\ref{jel_psi}) as
\begin{align}
	\left< j_{{\rm el},i} \right>^{\psi}_{\omega}
	=&
	\int {\rm d} \varepsilon
	\left( - \frac{\partial f}{\partial \varepsilon} \right)
	\varepsilon \sigma(\varepsilon) (- \partial_{i} \psi)_{\bm{Q}}
	\notag \\
	& -
	\int {\rm d} \varepsilon
	\left( - \frac{\partial f}{\partial \varepsilon} \right)
	\varepsilon D(\varepsilon) \nu(\varepsilon)
	\left( - \frac{\partial_{i} T}{T} \right)_{\bm{Q}}
	\notag \\
	=&
	\int {\rm d} \varepsilon
	\left( - \frac{\partial f}{\partial \varepsilon} \right)
	\varepsilon \sigma(\varepsilon)
		\left( - \partial_{i} \psi  - \frac{\partial_{i} T}{T} \right)_{\bm{Q}}  . 
	\label{rewrite_jel_psi}
\end{align}
 This shows the \lq\lq equivalence'' of the mechanical force $\partial_i \psi$ 
and the statistical force $\partial_i T/T$, and forms a basis of Luttinger's thermal linear-response theory.

\subsection{Magnon-drag process: Spin chemical potential}
\label{Spin chemical potential}

 Let us apply a similar procedure to the magnon-drag process. 
 For this purpose, we calculate the magnon-drag electron current 
in response to a spatially-modulated potential, 
$\psi \propto \psi_{\bm Q} \, e^{i({\bm Q} \cdot {\bm r} - \omega t)}$, 
with finite wave vector ${\bm Q}$. 
 As in the preceding subsection, we consider the diffusion-type vertex corrections 
and the diagrams in Fig.~\ref{fig:Kij} (b). 
 The result is obtained as 
\begin{align}
	\left< j_{{\rm el},i} \right>_{\rm drag}
	=&
	- \frac{1}{2 s_{0}}
	\frac{\sigma_{\uparrow} - \sigma_{\downarrow}}{e^2}
	 \left( 1	- \frac{\beta_{{\rm el}}}{\alpha} \right) T \mathcal{S}_{{\rm mag}} 
	\left( - \partial_{i} \psi \right)
	\notag \\
	&
	- \partial_{i} 
	\left( D_{\uparrow} \delta n^{\uparrow}_{\rm el} + D_{\downarrow} \delta n^{\downarrow}_{\rm el} \right),
	\label{discuss_jel_diffusive}
\end{align}
where
\begin{align}
	\delta n^{\sigma}_{\rm el}
	=&
	\frac{\sigma}{2 s_{0}} 
	\frac{\sigma_{\sigma}}{e^2}
	 \left( 1	- \frac{\beta_{{\rm el}}}{\alpha} \right) T \mathcal{S}_{{\rm mag}} 
	\frac{Q^{2}}{D_{\sigma}Q^{2} - i \omega} \psi_{\bm{Q}},
	\label{discuss_nels}
\end{align}
is the change of the electron density (of spin $\sigma$) caused by the perturbation 
$\psi_{\bm{Q}}$, 
and $D_{\sigma} = (v_{ {\rm F}\sigma}^{2}/3) \tau_{\sigma}$ is the diffusion constant. 
 From the form of Eq.~(\ref{discuss_nels}), 
it is natural to regard the density change $\delta n^{\sigma}_{\rm el}$ as caused by the change of the electrons' chemical potential, 
instead of temperature as in Eq.~(\ref{def_deltaT}). 
 Namely, Eq.~(\ref{discuss_nels}) in the \lq slow' limit, $\omega \to 0$, may be compared with 
\begin{align}
	\delta n^{\sigma}_{{\rm el}}
	&=
	\int {\rm d} \varepsilon \, \nu_{\sigma} (\varepsilon) 
	[ \, f(\varepsilon; T \mu + \delta \mu_{\sigma}) - f(\varepsilon; T,  \mu) \, ] 
	\notag \\
	& \simeq 
	\int {\rm d} \varepsilon \, \nu_{\sigma} (\varepsilon) 
	\left( - \frac{\partial f}{\partial \varepsilon} \right) \delta \mu_{\sigma} 
	\notag \\
	& \simeq 
	\nu_{\sigma} \delta \mu_{\sigma}  ,
	\label{discuss_def_mus}
\end{align}
where $\delta \mu_\sigma$ is the change in (spin-dependent) chemical potential. 
 From the comparison, we may identify \cite{com_deltaT} 
\begin{align}
	\delta \mu_{\sigma}
	&= 
	\frac{\sigma}{2 s_{0}}  
    \left( 1	- \frac{\beta_{\rm s}}{\alpha} \right) T \mathcal{S}_{{\rm mag}} 
	\psi_{\bm{Q}} , 
	\label{discuss_mus_psi}
\noindent \\
	&= 
	\frac{\sigma}{2 s_{0}}  
	\left( 1	- \frac{\beta_{\rm s}}{\alpha} \right) T \mathcal{S}_{\rm mag} 
	\left( - \frac{\delta T_{\bm{Q}}}{T} \right) . 
	\label{discuss_mus_T}
\end{align}
 In the second line, we used the Einstein-Luttinger relation, 
$\psi_{\bm{Q}} = - \delta T_{\bm{Q}} /T$. 
 Note that Eq.~(\ref{discuss_mus_T}) is consistent with Eq.~(\ref{eq:delta_mu_sigma}). 
 Using Eq.~(\ref{discuss_mus_T}) for $\delta \mu_\sigma$ in Eq.~(\ref{discuss_def_mus}), 
we rewrite Eq.~(\ref{discuss_jel_diffusive}) as
\begin{align}
	\left< j_{{\rm el},i} \right>_{\rm drag}
	=&
	- \frac{1}{2 s_{0}}
	\frac{\sigma_{\uparrow} - \sigma_{\downarrow}}{e^2}
	 \left( 1	- \frac{\beta_{{\rm el}}}{\alpha} \right) T \mathcal{S}_{{\rm mag}} 
	\left( - \partial_{i} \psi \right)
	\notag \\
	&
	+ \frac{1}{e^2} \sigma_{\uparrow} 
	\left( - \partial_{i} \delta \mu_{\uparrow} \right)
	+ \frac{1}{e^2} \sigma_{\downarrow} 
	\left( - \partial_{i} \delta \mu_{\downarrow} \right)
	\label{discuss_jel_diffusive_mu}
	\\
	=&
	- \frac{1}{2 s_{0}}
	\frac{\sigma_{\uparrow} - \sigma_{\downarrow}}{e^2}
	 \left( 1	- \frac{\beta_{{\rm el}}}{\alpha} \right) T \mathcal{S}_{{\rm mag}} 
	\notag \\
	& \times
	\left( - \partial_{i} \psi - \frac{\partial_{i} T}{T} \right) . 
\end{align}
 This reproduces the form of Eq.~(\ref{eq:def_jel_psi}).

 The nonequilibrium chemical potential $\delta \mu_{\sigma}$ is spin dependent, 
$\delta \mu_\uparrow = - \delta \mu_\downarrow$ 
(because of the overall factor $\sigma = \pm 1$). 
 Thus the electrons feel the effects of the nonequilibrium magnons 
as a \lq\lq spin chemical potential'', 
or spin accumulation, $\mu_{\rm s} = \mu_\uparrow - \mu_\downarrow$. 
 This is quite natural since the local change $\delta T$ in temperature modulates 
the magnon density, and the balance of the \lq\lq reaction''  
\begin{align}
 {\rm m} \  + \  e_\uparrow \  \rightleftarrows \  \, e_\downarrow , 
\label{eq:reaction}
\end{align}
shifts in the left or the right direction.  
 Here, m, $e_\uparrow$ and $e_\downarrow$ represent a magnon, an electron with spin up, 
and an electron with spin down, respectively. 
 If we focus on the electrons ($e_\uparrow$ and $e_\downarrow$), this is precisely 
controlled by the chemical-potential difference, $\mu_\uparrow - \mu_\downarrow$. 
 This process corresponds to the first term (the spin-transfer term). 
 The absence of the causality relationship, as discussed at the end of Sec.~IV, 
may be due to the local equilibrium nature of this process.

 The second term (proportional to $\beta_{\rm el}/\alpha$) acts in the opposite way; 
it increases the density of up-spin (down-spin) electrons in the hotter (colder) region.  
 Let us interpret this effect in terms of momentum transfer process. 
 For this, we consider the effects of magnon flow. 
 The magnons flow from the hotter to the colder region, and will scatter 
electrons into the colder region. 
 If a magnon is absorbed by an electron, the scattered electron has down spin and flows downstream. 
 This means that the down-spin electrons flow to colder regions and this effect will increase 
the density of down-spin electrons in the colder region. 
 There is also a reverse process: if a down-spin electron emits a magnon and flips its spin, 
and if the magnon flows downstream, the up electron will flow upstream. 
 This process will increase the density of up-spin electrons in the hotter region.

\section{summary}

 In this paper, we studied magnon-drag electron flow induced by a temperature gradient. 
 The analysis is based on a microscopic model that contains spin relaxation, 
and on the linear response theory due to Luttinger that exploits a gravitational potential $\psi$. 
 The obtained result is physically interpreted in terms of the spin-transfer process 
and the momentum-transfer process from the magnons to the electrons. 
 It is found that the effect of nonequilibrium magnons yields a nonzero spin chemical 
potential of the electrons. 
 In the process, we gave a microscopic procedure that leads to the Luttinger's form 
of the response, namely, a combination of the form, $-\partial_i \psi - \partial_i T/T$. 
 We supplemented the analysis with a phenomenological one that is based 
on the spin-motive force, 
and found that the agreement with the microscopic result is good for the 
dissipative $\beta$-term, but differs slightly for the Berry-phase (spin-transfer) term.

\section*{acknowledgement}

 We are grateful to Y.~Imai for fruitful discussions. 
 Valuable comments by G.~E.~W.~Bauer and J.~P.~Heremans are also appreciated. 
 This work is supported by JSPS KAKENHI Grant Numbers 25400339, 15H05702 and 17H02929. 
 TY is supported by a Program for Leading Graduate Schools 
``Integrative Graduate Education and Research in Green Natural Sciences''. 
RD is a member of the D-ITP consortium, a program of the Netherlands Organisation 
for Scientific Research (NWO) that is funded by the Dutch Ministry of Education, 
Culture and Science (OCW). This work is in part funded by the Stichting voor 
Fundamenteel Onderzoek der Materie (FOM) and the European Research Council (ERC).

\appendix

\section{Vertex corrections}\label{sec:vertexcorrection} 

In this Appendix, we calculate the vertex corrections to the electron spin $\sigma^{\pm}$ 
due to impurity potentials in the ladder approximation.
 The renormalized vertex $\Lambda^{\pm}_{\sigma \bar{\sigma}}$ satisfies 
\begin{align}
	(\Lambda^{\pm}_{\sigma \bar{\sigma}})^{ab}
	=
	(\sigma^{\pm})_{\sigma \bar{\sigma}}
	+ \Gamma_{0} Y^{ab}_{\sigma \bar\sigma} (\Lambda^{\pm}_{\sigma \bar{\sigma}})^{ab},
	\label{def_Lambdap}
\end{align}
where $(\sigma^{+})_{\uparrow \downarrow} = (\sigma^-)_{\downarrow \uparrow} = 1$ 
(other elements vanish), 
\begin{align}
	\Gamma_{0}
	= n_{\rm i} u_{\rm i}^{2} - n_{\rm s} u_{\rm s}^{2} \overline{S_{z}^{2}},
	\label{def_Gamma0}
\end{align}
and $Y^{ab}_{\sigma \bar\sigma} = \sum_{\bm{k}}G_{{\bm k} \sigma}^a G_{{\bm k} \bar{\sigma}}^b$ 
with $\bar{\sigma} = - \sigma$. 
 We write the Green's function as 
$G_{{\bm k} \sigma}^a = (i \varepsilon_{a} + \sigma M - \hbar^2{\bm k}^2/2m - \Sigma_\sigma^a)^{-1}$, 
where $a$, $b$ specify retarded (R) or advanced (A), 
namely, $a={\rm R}$ for $\varepsilon_a > 0$, and  $a={\rm A}$ for $\varepsilon_a < 0$. 
 Writing the self-energy as 
\begin{align}
	\Sigma_\sigma^a 
	= \Gamma_1 g_\sigma^a + \Gamma_2 g_{\bar\sigma}^b ,  
	\label{Sigma}
\end{align}
with  
$\Gamma_{1} = n_{\rm i} u_{\rm i}^{2} + n_{\rm s} u_{\rm s}^{2} \overline{S_{z}^{2}}$, 
$\Gamma_{2} = 2 n_{\rm s} u_{\rm s}^{2} \overline{S_{\perp}^{2}}$ [Eq.~(\ref{eq:Gamma})], 
and $g^{a}_{\sigma} = \sum_{\bm{k}} G_{{\bm k} \sigma}^a$, 
we evaluate $Y^{ab}_{\sigma \bar\sigma}$ as
\begin{align}
	Y^{ab}_{\sigma \bar\sigma}
&= 
\frac{g^{ab}_{\sigma \bar{\sigma}} }
           {i \varepsilon_{ba} - 2\sigma M + \Sigma^{ab}_{\sigma \bar{\sigma}}} ,
	\label{trans_Ys}
\end{align}
where $\varepsilon_{ba} = \varepsilon_b - \varepsilon_a$, 
$g^{ab}_{\sigma \sigma'} = g^{a}_{\sigma} - g^{b}_{\sigma'}$ 
and $\Sigma^{ab}_{\sigma \sigma'} = \Sigma^{a}_{\sigma} - \Sigma^{b}_{\sigma'}$. 
 Then, from Eq.~(\ref{def_Lambdap}), we obtain 
\begin{align}
	(\Lambda^{\pm}_{\sigma \bar{\sigma}})^{ab}
&= \frac{(\sigma^{\pm})_{\sigma \bar{\sigma}}}{1 - \Gamma_{0} Y^{ab}_{\sigma \bar\sigma} } 
 = \frac{i \varepsilon_{ba} - 2\sigma M + \Sigma^{ab}_{\sigma \bar\sigma} }
           {i \varepsilon_{ba} - 2\sigma M + \Delta^{ab}_{\sigma \bar\sigma} } 
    (\sigma^{\pm})_{\sigma \bar{\sigma}}  , 
	\label{solve_Lambdap}
\end{align}
where $\Delta^{ab}_{\sigma \sigma'} = \Delta^a_{\sigma} - \Delta^b_{\sigma'}$ with 
\begin{align}
	\Delta^a_\sigma
&=
	(\Gamma_{1} - \Gamma_{0}) g^a_\sigma + \Gamma_{2} g^a_{\bar\sigma} 
\nonumber \\
&= 2 n_{\rm s} u_{\rm s}^{2} 
    ( \overline{S_{z}^{2}} g^a_\sigma + \overline{S_{\perp}^{2}} g^a_{\bar\sigma} )  . 
	\label{def_Delta}
\end{align}
 Explicitly, $\Lambda^+$ and $\Lambda^-$ are given by 
\begin{align}
  (\Lambda^{+}_{\uparrow \downarrow})^{ab}
&= \frac{i \varepsilon_{ba} - 2M + \Sigma^{ab}_{\uparrow \downarrow} }
           {i \varepsilon_{ba} -2M + \Delta^{ab}_{\uparrow \downarrow} } , 
	\label{result_Lambdap}
\\
  (\Lambda^{-}_{\downarrow \uparrow})^{ab}
&= \frac{i \varepsilon_{ba} + 2M + \Sigma^{ab}_{\downarrow \uparrow} }
           {i \varepsilon_{ba} +2M + {\Delta}^{ab}_{\downarrow \uparrow} } . 
	\label{result_Lambdam}
\end{align}
(Other elements vanish, $\Lambda^{+}_{\downarrow \uparrow}  = \Lambda^{-}_{\uparrow \downarrow}  = 0$, etc.)  
Therefore,
\begin{align}
 G_{{\bm k} \uparrow}^a (\Lambda^{+}_{\uparrow \downarrow})^{ab} G_{{\bm k} \downarrow}^b 
&= \frac{G_{{\bm k} \uparrow}^a - G_{{\bm k} \downarrow}^b}{i \varepsilon_{ba} - 2M + \Delta^{ab}_{\uparrow \downarrow} } , 
	\label{result_Lambdap_GG}
\\ 
 G_{{\bm k} \downarrow}^a (\Lambda^{-}_{\downarrow \uparrow})^{ab} G_{{\bm k} \uparrow}^b 
&= \frac{G_{{\bm k} \downarrow}^a - G_{{\bm k} \uparrow}^b}{i \varepsilon_{ba} + 2M + {\Delta}^{ab}_{\downarrow \uparrow} } .
	\label{result_Lambdam_GG}
\end{align}
For example,
\begin{align}
	G^{\rm R}_{{\bm k} \uparrow} (\Lambda^{+}_{\uparrow \downarrow})^{\rm RA} G^{\rm A}_{{\bm k} \downarrow}
&= 
	\frac{G^{\rm R}_{{\bm k} \uparrow} - G^{\rm A}_{{\bm k} \downarrow}}{-2M + \Delta^{\rm RA}_{\uparrow \downarrow} }
\notag \\
& \simeq 
	- \frac{1}{2M} \left( 1 + \frac{\Delta^{\rm RA}_{\uparrow \downarrow} }{2M} \right)
	\left( G^{\rm R}_{{\bm k} \uparrow} - G^{\rm A}_{{\bm k} \downarrow} \right),
	\label{eq:example_Lambdap}
\\ 
	G^{\rm R}_{{\bm k} \downarrow} (\Lambda^{-}_{\downarrow \uparrow})^{\rm RA} G^{\rm A}_{{\bm k} \uparrow}
&= 
	\frac{G^{\rm R}_{{\bm k} \downarrow} - G^{\rm A}_{{\bm k} \uparrow}}{2M + {\Delta}^{\rm RA}_{\downarrow \uparrow} }
\notag \\
& \simeq 
	\frac{1}{2M}
	\left( 1 - \frac{{\Delta}^{\rm RA}_{\downarrow \uparrow}}{2M} \right)
	\left( G^{\rm R}_{{\bm k} \downarrow} - G^{\rm A}_{{\bm k} \uparrow} \right) . 
	\label{eq:example_Lambdam}
\end{align}
 In the second lines, we assumed that $\Delta^{ab}$'s, 
which are on the order of spin relaxation rate, are much smaller than the exchange 
splitting $M$.

\section{Details of microscopic calculation}\label{sec:micro_cal}

 In this Appendix, we present the details of the calculation of the magnon-drag electron current. 
 It is divided into the electron part and the magnon part.

\subsection{Electron part}

 As described in the text, the electron part, given by Eq.~(\ref{def_Ei}), 
contributes in two different ways, 
${\cal E}^{(2)}_i \equiv {\cal E}_i ({\bm q}, \nu - i\eta; \omega + 2i\eta)$
and ${\cal E}^{(1)}_i \equiv {\cal E}_i ({\bm q}, \nu + i\eta; \omega + i\eta)$, 
where $\eta$ is a positive infinitesimal. 
 For the magnon-drag contribution, 
the former is calculated by setting $\omega=0$ and retaining the $\nu$-linear terms, 
and the latter by setting $\nu=0$ and retaining the $\omega$-linear terms. 
 They are given, respectively, by 
\begin{align}
	& \mathcal{E}^{(2)}_{i}
	=
	\frac{\nu}{2 \pi} q_{\ell}
	\left( \mathcal{E}'_{i \ell} + \mathcal{E}''_{i \ell} \right) ,
	\label{def_Ei_omega0}
\end{align}
with
\begin{widetext}
\begin{align}
	\mathcal{E}'_{i \ell}
	&= 	
	-i \sum_{\bm{k}}
	v_{i} v_\ell 
	\left[ 
		 G^{\rm R}_{\downarrow}
		\left\{ 
			(\Lambda^{-}_{\downarrow \uparrow})^{\rm RR}
			\left( G^{\rm R}_{\uparrow} \right)^{2}
			(\Lambda^{+}_{\uparrow \downarrow})^{\rm RA}
			- 
			(\Lambda^{-}_{\downarrow \uparrow})^{\rm RA}
			\left( G^{\rm A}_{\uparrow} \right)^{2}
			(\Lambda^{+}_{\uparrow \downarrow})^{\rm AA}
		\right\}
		G^{\rm A}_{\downarrow}
	\right. 
	\notag \\
	& \hskip 20mm 
	\left.
		 + \, 
		G^{\rm R}_{\uparrow}
		\left\{ 
			(\Lambda^{+}_{\uparrow \downarrow})^{\rm RR}
			\left( G^{\rm R}_{\downarrow} \right)^{2}
			(\Lambda^{-}_{\downarrow \uparrow})^{\rm RA}
          - 
			(\Lambda^{+}_{\downarrow \uparrow})^{\rm RA}
			\left( G^{\rm A}_{\downarrow} \right)^{2}
			(\Lambda^{-}_{\downarrow \uparrow})^{\rm AA}
		\right\}
		G^{\rm A}_{\uparrow}
	\right] , 
	\label{def_E1iell}
	\\
	\mathcal{E}''_{i \ell}
	&= 	
	- \sum_{\bm{k}}
	v_{i} v_\ell 
	\Im
	\left[ 
		G^{\rm R}_{\downarrow}
		(\Lambda^{-}_{\downarrow \uparrow})^{\rm RR}
		\left( G^{\rm R}_{\uparrow} \right)^{2}
		(\Lambda^{+}_{\uparrow \downarrow})^{\rm RR}
		G^{\rm R}_{\downarrow}
		+
		G^{\rm R}_{\uparrow}
		(\Lambda^{+}_{\uparrow \downarrow})^{\rm RR}
		\left( G^{\rm R}_{\downarrow} \right)^{2}
		(\Lambda^{-}_{\downarrow \uparrow})^{\rm RR}
		G^{\rm R}_{\uparrow}
	\right],
	\label{def_E2iell}
\end{align}
and
\begin{align}
	\mathcal{E}^{(1)}_{i}
	& \simeq 
 \frac{1}{2 \pi} q_{\ell}
	\sum_{\bm{k}}
	v_{i} v_{\ell}
	\left[ 
		- G^{\rm R}_{\downarrow}
		(\Lambda^{-}_{\downarrow \uparrow})^{\rm RA}
		\left( G^{\rm A}_{\uparrow} \right)^{2}
		(\Lambda^{+}_{\uparrow \downarrow})^{\rm AA}
		G^{\rm A}_{\downarrow}
		+
		G^{\rm R}_{\uparrow}
		(\Lambda^{+}_{\uparrow \downarrow})^{\rm RR}
		\left( G^{\rm R}_{\downarrow} \right)^{2}
		(\Lambda^{-}_{\downarrow \uparrow})^{\rm RA}
		G^{\rm A}_{\uparrow}
	\right] , 
\end{align}
where $G^{\rm R(A)}_{\sigma} = G^{\rm R(A)}_{\bm{k} \sigma} (0)$.

To calculate $\mathcal{E}'_{i \ell}$, 
we use Eqs.~(\ref{result_Lambdap_GG})-(\ref{result_Lambdam_GG}) and the approximations 
as in Eqs.~(\ref{eq:example_Lambdap})-(\ref{eq:example_Lambdam}) 
valid for weak spin relaxation (compared to $M$). 
 With short notations, $\Delta^{ab} = \Delta^{ab}_{\uparrow\downarrow}$ and 
$\tilde \Delta^{ab} = \Delta^{ab}_{\downarrow\uparrow}$, we write 
\begin{align}
	\mathcal{E}'_{i \ell}
	\simeq & 
	\frac{i}{(2M)^{2}}
	\sum_{\bm{k}} v_{i} v_{\ell}
	\left[ 
		\left( 1 + \frac{\Delta^{\rm RA} - \tilde{\Delta}^{\rm RR}}{2M} \right)
		(G^{\rm R}_{\downarrow} - G^{\rm R}_{\uparrow})
		(G^{\rm R}_{\uparrow} - G^{\rm A}_{\downarrow})
		-
		\left( 1 + \frac{\Delta^{\rm AA} - \tilde{\Delta}^{\rm RA}}{2M} \right)
		(G^{\rm R}_{\downarrow} - G^{\rm A}_{\uparrow})
		(G^{\rm A}_{\uparrow} - G^{\rm A}_{\downarrow})
	\right.
	\notag \\
	& \qquad \qquad \qquad 
	\left.
		-
		\left( 1 + \frac{\Delta^{\rm RA} - \tilde{\Delta}^{\rm AA}}{2M} \right)
		(G^{\rm R}_{\uparrow} - G^{\rm A}_{\downarrow})
		(G^{\rm A}_{\downarrow} - G^{\rm A}_{\uparrow})
		+
		\left( 1 + \frac{\Delta^{\rm RR} - \tilde{\Delta}^{\rm RA}}{2M} \right)
		(G^{\rm R}_{\uparrow} - G^{\rm R}_{\downarrow})
		(G^{\rm R}_{\downarrow} - G^{\rm A}_{\uparrow})
	\right]
	\notag \\
	\simeq & 
	\frac{i}{(2M)^{2}}
	\sum_{\bm{k}} v_{i} v_{\ell}
	\left[ 
		\frac{\Delta_{1}}{2M} G^{\rm R}_{\uparrow} G^{\rm A}_{\uparrow}
		- \frac{\Delta_{2}}{2M} G^{\rm R}_{\downarrow} G^{\rm A}_{\downarrow}
		- 2 i \Im \left\{ (G^{\rm R}_{\uparrow} - G^{\rm R}_{\downarrow})^{2} \right\}
	\right],
	\label{expand_E1}
\end{align}
\end{widetext}
where 
$\Delta_{1} 
\equiv \Delta^{\rm RA} - \Delta^{\rm RR} + \tilde{\Delta}^{\rm RA} - \tilde{\Delta}^{\rm AA}$
and
$\Delta_{2}
\equiv \Delta^{\rm RA} - \Delta^{\rm AA} + \tilde{\Delta}^{\rm RA} - \tilde{\Delta}^{\rm RR}$,
and we retained the leading terms with respect to the electron damping.
On the other hand, $\mathcal{E}''_{i \ell}$ is calculated as
\begin{align}
	\mathcal{E}''_{i \ell}
	\simeq
	- \frac{2}{(2M)^{2}}
	\sum_{\bm{k}} v_{i} v_{\ell} 
	\Im \left[ (G^{\rm R}_{\uparrow} - G^{\rm R}_{\downarrow})^{2} \right] . 
	\label{trans_E2}
\end{align}
 Therefore, we have  
\begin{align}
	\mathcal{E}'_{i \ell} + \mathcal{E}''_{i \ell}
&=
	\frac{i}{(2M)^{2}}
	\sum_{\bm{k}} v_{i} v_{\ell}
	\left[ 
		\frac{\Delta_{1}}{2M} G^{\rm R}_{\uparrow} G^{\rm A}_{\uparrow}
		- \frac{\Delta_{2}}{2M} G^{\rm R}_{\downarrow} G^{\rm A}_{\downarrow}
	\right] 
\nonumber \\
&=
	\frac{i \delta_{i \ell}}{(2M)^{2}}
	\frac{2 \pi}{e^{2}}
	\left[ 
		\frac{\Delta_{1}}{2M} \sigma_{\uparrow}
		- \frac{\Delta_{2}}{2M} \sigma_{\downarrow}
	\right] . 
	\label{result_E1+E2}
\end{align}
 Here we noted
\begin{align}
	\sum_{\bm{k}}
	v_{i} v_{\ell} G^{\rm R}_{\sigma} G^{\rm A}_{\sigma}
	=
	\delta_{i \ell} \frac{v_{ {\rm F}\sigma}^{2}}{3}
	\frac{\pi \nu_{\sigma}}{\gamma_{\sigma}}
	=
	\delta_{i \ell} \frac{2 \pi}{e^{2}} \sigma_{\sigma},
	\label{int_vvRsAs}
\end{align}
with $\sigma_{\sigma} = e^{2} (v_{ {\rm F}\sigma}^{2}/3) \tau_{\sigma}$
being the conductivity of spin-$\sigma$ electrons. 
 From Eq.~(\ref{def_Delta}), we have 
$\Delta_{1} = 2\Delta^{\rm RA}_{\downarrow\downarrow}$ and $\Delta_{2} = 2\Delta^{\rm RA}_{\uparrow\uparrow}$ with 
$\Delta^{\rm RA}_{\sigma \sigma} 
= - 4\pi i n_{\rm s} u_{\rm s}^{2} (\overline{S^{2}_{z}}  \nu_{\sigma} + \overline{S^{2}_{\perp}} \nu_{\bar\sigma}) $, 
and thus
\begin{align}
	\frac{\Delta_{1}}{2M} \sigma_{\uparrow}
	- \frac{\Delta_{2}}{2M} \sigma_{\downarrow}
	=
	- 2i \beta_{\rm el}
	\left( \sigma_{\uparrow} - \sigma_{\downarrow} \right) , 
	\label{rewrite_D1D2}
\end{align}
where $\beta_{\rm el}$ is given by Eq.~(\ref{def_beta}). 
 Using these relations in Eq.~(\ref{def_Ei_omega0}), we obtain 
\begin{align}
	\mathcal{E}^{(2)}_{i}
	=
	\beta_{\rm el} \frac{\nu}{2M^{2}} q_{i}  
	\frac{ \sigma_{\uparrow} - \sigma_{\downarrow} }{e^2} .
	\label{result_Ei_omega0_app}
\end{align}
 The $\omega$-linear terms in $\mathcal{E}^{(2)}_{i}$ [as given in Eq.~(\ref{result_Ei_2})], 
which contributes to the spin-motive forces, can be obtained in a similar way.

 Similarly, we obtain 
\begin{align}
	\mathcal{E}^{(1)}_{i}
	& \simeq  
	\frac{1}{2 \pi} \frac{1}{(2M)^{2}} q_{\ell}
	\sum_{\bm{k}}
	v_{i} v_{\ell}
	\left[ 
		G^{\rm R}_{\uparrow} G^{\rm A}_{\uparrow}
		- G^{\rm R}_{\downarrow} G^{\rm A}_{\downarrow}
	\right]
	\notag \\
	& \simeq  
	\frac{1}{(2M)^2} q_i \frac{\sigma_{\uparrow} - \sigma_{\downarrow}}{e^{2}} . 
	\label{result_Ei_nu0_app}
\end{align}

\subsection{Magnon part}

 For the magnon part, we encounter the following integrals, 
\begin{align}
	I_1 
&= \frac{1}{2\pi} \int {\rm d} \nu \left( - \frac{\partial n}{\partial \nu} \right) \nu^{2}
	\sum_{\bm{q}} u_{i} q_{j} D^{\rm R}_{\bm{q}} (\nu)  D^{\rm A}_{\bm{q}} (\nu) , 
\\ 
	I_2 
&= \frac{1}{\pi} \int {\rm d} \nu \, n(\nu) \, \nu
	\sum_{\bm{q}} u_{i} q_{j} \Im \left[ \left( D^{\rm R}_{\bm{q}} (\nu) \right)^{2} \right] ,
\\
	I_3 
&= \frac{1}{\pi} \int {\rm d} \nu n(\nu) \, \sum_{\bm{q}} u_{i} q_{j}
	\Im \left[ D^{\rm R}_{\bm{q}} \right] ,
\end{align}
 To calculate $I_1$, we use 
$D^{\rm R}_{\bm q} (\nu) D^{\rm A}_{\bm q} (\nu) 
\simeq (\pi /\alpha \nu) \, \delta (\nu - \omega_{\bm q})$. 
 Then,  
\begin{align}
	I_1 
	& \simeq 
	\frac{1}{2 \alpha}
	\sum_{\bm{q}}
	\omega_{\bm{q}} u_{i} q_{j}
	\left( - \frac{\partial n}{\partial \nu} \right)_{\nu = \omega_{\bm{q}}}
	\notag \\
	& =
	\frac{1}{2 \alpha}
	T \frac{\partial}{\partial T}
	\sum_{\bm{q}}
	n(\omega_{\bm{q}}) u_{i} q_{j} . 
\label{eq:I1_a}
\end{align}
 By noting 
$ (\partial/\partial q_i) k_{\rm B} T 
   \ln \left( 1 - e^{- \hbar \omega_{\bm{q}} /k_{\rm B} T} \right) 
 = n(\omega_{\bm{q}}) u_i$, we see 
\begin{align}
	\sum_{\bm{q}} n(\omega_{\bm{q}}) u_{i} q_{j}
    &= 
	\sum_{\bm{q}} q_j \frac{\partial}{\partial q_i} 
	 k_{\rm B} T \ln \left( 1 - e^{- \hbar \omega_{\bm{q}} / k_{\rm B} T} \right)  
	\notag \\
	& = - \delta_{ij} \Omega_{\rm mag}  , 
\label{eq:Omega_mag_identity_1}
\end{align}
where
\begin{align}
 \Omega_{\rm mag} &= k_{\rm B} T \sum_{\bm{q}}  
  \ln \left( 1 - e^{- \hbar \omega_{\bm{q}}/k_{\rm B} T } \right) , 
\label{eq:Omega_mag}
\end{align}
is the thermodynamic potential of magnons.  Therefore, 
\begin{align}
  I_1 &= 
	- \frac{1}{2 \alpha} \delta_{ij} 
	T \frac{\partial}{\partial T} \Omega_{\rm mag} 
 =
	\frac{1}{2\alpha} \delta_{ij}
	T \mathcal{S}_{\rm mag},
	\label{int_magnon_RA_app}  
\end{align}
where 
$\mathcal{S}_{\rm mag} = - \partial \Omega_{\rm mag}/\partial T$ is the entropy (density) 
of magnons.

 For $I_2$, we use 
$u_i (D^{\rm R})^2 = \partial D^{\rm R}/\partial q_i$ and 
$\Im D^{\rm R}_{\bm q} (\nu) \simeq - \pi \delta (\nu - \omega_{\bm q})$, 
and calculate as 
\begin{align}
	I_2 
	&=
	- \frac{\delta_{ij}}{\pi}
	\int {\rm d} \nu \, n (\nu) \, \nu
	\sum_{\bm{q}} \Im \left[ D^{\rm R}_{\bm{q}} \right]
	\notag \\
	&\simeq
	\delta_{ij} \sum_{\bm{q}} \omega_{\bm{q}} n(\omega_{\bm{q}})
	\notag \\
	&=
	\delta_{ij} \mathcal{E}_{\rm mag} . 
	\label{int_magnon_RR_app}
\end{align}

 Similarly, $I_3$ is calculated as 
\begin{align}
	I_3 
&\simeq -  \sum_{\bm{q}} u_{i} q_{j} n(\omega_{\bm q})
= \delta_{ij} \Omega_{\rm mag} . 
\end{align}

\section{Semi-classical analysis based on spin-motive force}\label{sec:pheno_anal}

 In this Appendix, we calculate 
\begin{align}
 \langle F_i \rangle &=  \frac{1}{s_0} \left\{ 
	{\rm Im} \langle \dot a^\dagger \partial_i a \rangle
	- \beta \, {\rm Re} \langle \dot a^\dagger \partial_i a \rangle \right\} , 
\label{eq:F_average}
\end{align}
semi-classically 
using the stochastic Landau-Lifshitz-Gilbert (LLG) equation.  
 This method has been used in the calculation of magnonic spin torques \cite{Kovalev2014,Kim2015}.

\subsection{Formulation}

 The stochastic LLG equation is given by 
\begin{align}
	\dot{\bm{n}}
	=
	- J \bm{n} \times \partial_{i}^{2} \bm{n}
	+ \bm{n} \times \bm{h}
	- \alpha \bm{n} \times \dot{\bm{n}} , 
	\label{stochasticLLG}
\end{align}
where ${\bm n}$ is the magnetization unit vector, 
and ${\bm h}$ is the Langevin noise field that satisfies the fluctuation-dissipation 
theorem, 
\begin{align}
	\left< h_{i}(\bm{r},t) h_{j}(\bm{r}',t') \right>
	= 2 \alpha s_{0} T  \delta_{ij} \delta (\bm{r} - \bm{r}') \delta (t - t'),
	\label{FDT}
\end{align}
where $T$ is the temperature. 
 We consider the case that the temperature is nonuniform and assume $T$ 
in Eq.~(\ref{FDT}) is position-dependent, $T = T(\bm{r})$, 
and calculate $\left< F_{i} \right>$ that is proportional to $\partial_i T$.

 In the complex notation, $a = (s_{0}/2)^{1/2}( \delta n_{x} + i \delta n_{y})$ 
and $h= h_{x} + i h_{y}$, Eq.~(\ref{stochasticLLG}) becomes 
\begin{align}
	i \dot{a}
	= (- J \partial_{i}^{2} + \Delta ) a + \alpha \dot{a} - \frac{1}{\sqrt{2 s_{0}}} \, h(\bm{r},t) , 
	\label{stochasticLLG_complex}
\end{align}
where $\Delta$ is the magnon energy gap, and $h$ satisfies 
\begin{align}
	\left< h(\bm{r},t) h^{*}(\bm{r}', t') \right>
	= 4 \alpha s_{0} T(\bm{r}) \delta(\bm{r} - \bm{r}') \delta (t - t').
	\label{FDT_complex}
\end{align}
 Using the retarded Green's function $D^{\rm R} $ that satisfies 
\begin{align}
	\left[ i \partial_{t} + J \partial_{i}^{2} - \Delta - \alpha \partial_{t} \right] D^{\rm R} 
	= \delta (\bm{r} - \bm{r}') \delta (t - t') , 
	\label{def_Green_LLG}
\end{align}
 Eq.~(\ref{stochasticLLG_complex}) is solved as 
\begin{align}
	a (\bm{r},t)
	= - \frac{1}{\sqrt{2 s_{0}}}
	\int {\rm d}t' \int {\rm d}\bm{r}' D^{\rm R} (\bm{r} - \bm{r}', t - t') h (\bm{r}', t') . 
	\label{solution_LLG}
\end{align}
 In the Fourier representation,  
$D^{\rm R}_{\bm{q}} (\nu) = (\nu - \omega_{\bm{q}} + i \alpha \nu)^{-1}$, 
where $\omega_{\bm{q}} = J q^{2} + \Delta$, it reads
\begin{align}
	a_{\bm{q}}(\nu)
	= - \frac{1}{\sqrt{2 s_{0}}} D^{\rm R}_{\bm{q}}(\nu) h (\bm{q}, \nu) , 
	\label{solution_LLG_Fourier}
\end{align}
and $a^*$ is given by the complex conjugate of Eq.~(\ref{solution_LLG}).

 For a quantum system (in the present case, magnons), 
we consider the Fourier transform of Eq.~(\ref{FDT}) with respect to time, 
wherein the temperature is replaced as $T \to \frac{\nu}{2} \coth \frac{\nu}{2T} = \nu [n(\nu)+\frac{1}{2}]$ 
for the Fourier component of frequency $\nu$. 
 Its gradient is thus replaced as 
\begin{align}
	\partial_i  T 
\ \to \  
  \nu \left( \frac{\partial n}{\partial T} \right) \partial_i T . 
	\label{derivative_coth}
\end{align}

\subsection{Calculation of \texorpdfstring{$\left< F_{i} \right>$}{<Fi>}}

 To obtain $\left< F_{i} \right>$, it is sufficient to calculate 
$\left< \dot{a}^{\dagger} \partial_{i} a \right>$. 
 With Eq.~(\ref{solution_LLG_Fourier}), this proceeds as follows, 
\begin{widetext}
\begin{align}
	\left< \dot{a}^{\dagger}(\bm{r},t) \partial_{i} a(\bm{r},t) \right>
	&= 
	\left< 
		\partial_{t}
		\left( - \frac{1}{\sqrt{2 s_{0}}} 
		\int {\rm d} t_{1} \int {\rm d} {\bm{r}_{1}}
		D^{\rm R}(\bm{r}-\bm{r}_{1}, t-t_{1})
		h (\bm{r}_{1},t_{1}) \right)^* 
	\right.
	\notag \\
	& \quad \ 
	\left.
		\times \, 
		\partial_{i}
		\left( - \frac{1}{\sqrt{2 s_{0}}} 
		\int {\rm d}t_{2} \int {\rm d} \bm{r}_{2}
		D^{\rm R} (\bm{r}-\bm{r}_{2},t - t_{2})
		h (\bm{r}_{2},t_{2})  \right)
	\right>
	\notag \\
	&= 
	\frac{1}{2 s_{0}}
	\iint {\rm d}t_{1} {\rm d}t_{2}
	\iint {\rm d}\bm{r}_{1} {\rm d} \bm{r}_{2}
	\left( \partial_{t} D^{\rm R} (\bm{r}-\bm{r}_{1}, t-t_{1}) \right)^* 
	\left( \partial_{i} D^{\rm R} (\bm{r}-\bm{r}_{2}, t-t_{2}) \right)
	\left< h^{*}(\bm{r}_{1},t_{1}) h(\bm{r}_{2},t_{2}) \right>
	\notag \\
	&= 
	\frac{4 \alpha s_{0}}{2 s_{0}}
	\int {\rm d} t_{1}
	\int {\rm d} \bm{r}_{1}
	\left( \partial_{t} D^{\rm R} (\bm{r}-\bm{r}_{1}, t-t_{1}) \right)^* 
	\left( \partial_{i} D^{\rm R} (\bm{r}-\bm{r}_{1}, t-t_{1}) \right)
	T (\bm{r}_{1})
	\notag \\
	&= 
	2 \alpha 
	\int {\rm d} t_{1}
	\int {\rm d} \bm{r}_{1}
	\sum_{\bm{q},\bm{q}',\bm{q}_{1}}
	\iint \frac{ {\rm d}\nu {\rm d}\nu'}{(2 \pi)^{2}}
	i \nu' i q_{i}
	D^{\rm A}_{\bm{q}'}(\nu') D^{\rm R}_{\bm{q}}(\nu) \, 
	T_{\bm{q}_{1}}
	{\rm e}^{i (\bm{q}'-\bm{q}+\bm{q}_{1})\cdot \bm{r}_{1}}
	{\rm e}^{- i (\nu - \nu')(t - t_{1})}
	{\rm e}^{i (\bm{q} - \bm{q}')\cdot \bm{r}}
	\notag \\
	&= 
	2 \alpha 
	\sum_{\bm{q},\bm{q}_{1}}
	\int \frac{ {\rm d}\nu}{2 \pi}
	i \nu \cdot 
	i \left( q_{i} + \frac{q_{1,i}}{2} \right)
	D^{\rm R}_{\bm{q}+\bm{q}_{1}/2} (\nu)
	D^{\rm A}_{\bm{q}-\bm{q}_{1}/2} (\nu) \, 
	T_{\bm{q}_{1}} {\rm e}^{i \bm{q}_{1}\cdot \bm{r}} , 
	\label{trans_F_Langevin}
\end{align}
where 
$D^{\rm A}_{\bm{q}}(\nu)
 \equiv \left( D^{\rm R}_{\bm{q}}(\nu) \right)^{*}
 = (\nu - \omega_{\bm{q}} - i \alpha \nu)^{-1}$. 
We are interested in the term linear in $q_{1}$, 
which, combined with $T_{\bm{q}_{1}}$, gives the temperature gradient. 
 Thus, 
\begin{align}
	\left< \dot{a}^{\dagger}(\bm{r},t) \partial_{i} a(\bm{r},t) \right>
	& \simeq  
	\alpha 
	\sum_{\bm{q},\bm{q}_{1}}
	\int \frac{ {\rm d}\nu}{2\pi}
	i \nu 
	D^{\rm R} D^{\rm A}
	i q_{1,i} T_{\bm{q}_{1}} {\rm e}^{i \bm{q}_{1}\cdot \bm{r}}
	+ 2 \alpha 
	\sum_{\bm{q},\bm{q}_{1}}
	\int \frac{ {\rm d}\nu}{2 \pi}
	i \nu i q_{i} u_{j} \, 
	2i \Im
	\left[
		\left( D^{\rm R} \right)^{2} D^{\rm A}
	\right]
    \frac{q_{1,j}}{2} T_{\bm{q}_{1}} {\rm e}^{i \bm{q}_{1} \cdot \bm{r}} , 
	\label{trans_F_Langevin_02}
\end{align}
where $D^{\rm R} = D^{\rm R}_{\bm q} (\nu)$ and $D^{\rm A} = D^{\rm A}_{\bm q} (\nu)$. 
 With the replacement (\ref{derivative_coth}), we obtain 
\begin{align}
	\left< \dot{a}^{\dagger}(\bm{r},t) \partial_{i} a(\bm{r},t) \right>
	& \simeq 
	\alpha 
	\left( \partial_{i} T \right)
	\frac{\partial}{\partial T}
    \left\{ 
	i \sum_{\bm{q}}
	\int \frac{ {\rm d}\nu}{2 \pi}
	\nu^{2} n(\nu)
	D^{\rm R} D^{\rm A}
	 - 	2 \sum_{\bm{q}}
	\int \frac{ {\rm d}\nu}{2 \pi}
	q_{i} u_{j} \nu^{2} n(\nu)
	\Im 
	\left[
		\left( D^{\rm R} \right)^{2} D^{\rm A}
	\right] 
    \right\} .
	\label{trans_F_Langevin_03}
\end{align}
\end{widetext}
 Using the relations, 
\begin{align}
	\sum_{\bm{q}}
	\int \frac{ {\rm d}\nu}{2 \pi}
	\nu^{2} n(\nu)
	D^{\rm R} D^{\rm A}
	& \simeq 
	\sum_{\bm{q}}
	\int \frac{ {\rm d}\nu}{2 \pi}
	\nu^{2} n(\nu)
	\cdot
	\frac{\pi}{\alpha \nu}
	\delta (\nu - \omega_{\bm{q}}) 
	\notag \\
   & = 
	\frac{1}{2 \alpha} \mathcal{E}_{\rm mag},
	\label{int_nu2nDRDA}
\end{align}
and
\begin{align}
	& \ \ \  
    \sum_{\bm{q}}
	\int \frac{ {\rm d}\nu}{2 \pi}
	\nu^{2} n(\nu)
	q_{i} u_{j}
	\Im
	\left[
		\left( D^{\rm R} \right)^{2} D^{\rm A}
	\right]
	\notag \\
	&= 
	\sum_{\bm{q}}
	\int \frac{ {\rm d}\nu}{2 \pi}
	\nu^{2} n(\nu)
	q_{i} u_{j}
	\Im
	\left[
		D^{\rm R}
		\frac{1}{2 i \alpha \nu}
		\left( D^{\rm A} - D^{\rm R} \right)
	\right]
	\notag \\
	& \simeq  
	\sum_{\bm{q}}
	\int \frac{ {\rm d}\nu}{2 \pi}
	\nu^{2} n(\nu)
	q_{i} u_{j}
	\Im
	\left[		
		\frac{1}{2 i \alpha \nu}
		D^{\rm R} D^{\rm A}
	\right]
	\notag \\
	& \simeq  - \frac{1}{4 \alpha^2} \sum_{\bm q} q_i u_j  n(\omega_{\bm q})
	\notag \\
    &=   \frac{1}{4 \alpha^{2}} \, \Omega_{\rm mag} \delta_{ij} , 
	\label{int_nu2nquDR2DA}
\end{align}
where $\Omega_{\rm mag} $ is given by Eq.~(\ref{eq:Omega_mag}), 
we obtain
\begin{align}
	\left< \dot{a}^{\dagger} \partial_{i} a \right>
	&= 
	\frac{1}{2}
	\left( \frac{\partial_{i}T}{T} \right)
	T \frac{\partial}{\partial T}
	\left[ i \mathcal{E}_{\rm mag} - \frac{1}{\alpha} \Omega_{\rm mag} \right]
	\notag \\
	&= 
	\frac{1}{2}
	\left( \frac{\partial_{i}T}{T} \right)
	\left[ 
		i T \frac{\partial}{\partial T} \mathcal{E}_{{\rm mag}}
		+ \frac{1}{\alpha} T \mathcal{S}_{{\rm mag}}
	\right]  . 
	\label{result_F_complex}
\end{align}
 From Eq.~(\ref{eq:F_average}), this leads to 
\begin{align}
 \langle F_i \rangle 
&=  \frac{1}{2 s_0} \left\{ - \frac{\partial {\cal E}_{\rm mag}}{\partial T}  
	+ \frac{\beta}{\alpha}  {\cal S}_{\rm mag} \right\} (- \partial_i T)  . 
\label{eq:phen_result_app}
\end{align}

\subsection{Comparison with the previous study}

 To compare the phenomenological result (\ref{eq:phen_result_app}) obtained here 
with the one obtained previously \cite{Flebus2016}, let us consider the case, $T \gg \Delta$, 
where every quantity shows power-law dependence on temperature $T$. 
 In this case, 
$T (\partial \mathcal{E}_{{\rm mag}}/ \partial T)  \simeq (1 + d/2) \mathcal{E}_{\rm mag}$ 
and $T \mathcal{S}_{{\rm mag}} \simeq (1 + 2/d) \mathcal{E}_{\rm mag}$, 
and Eq.~(\ref{eq:phen_result_app}) becomes 
\begin{align}
	\left< F_i \right> 
	&= 
	- \frac{1}{2 s_{0}}  
	\left( 1 + \frac{d}{2} \right) 
    \left( 1 - \frac{2}{d} \frac{\beta}{\alpha} \right) \mathcal{E}_{{\rm mag}}  
    \left( - \frac{\partial_{i} T}{T} \right) . 
	\label{result_jsmf_Langevin_smallD_expand}
\end{align}
 Compared with the result of Ref.~\cite{Flebus2016},
the coefficient of $\beta/\alpha$ is different by a factor of 2.


\begin{thebibliography}{99}
\bibitem{com1}
 It is possible to argue this differently if we note  
\begin{align}
 (\sigma_{\uparrow} - \sigma_{\downarrow}) \beta_{\rm el}
 =  (\sigma_{\uparrow} - \sigma_{\downarrow}) \beta_{\rm s}
   +  (\sigma_{\uparrow} + \sigma_{\downarrow}) \beta_{\rm c} , 
\nonumber
\end{align}
where 
\begin{align}
 \beta_{\rm s} &= \frac{\pi n_{\rm s} u_{\rm s}^{2}}{M}
	            ( \overline{S^{2}_{\perp}} + \overline{S^{2}_{z}} ) \, \nu_{+} , 
\nonumber
\\
 \beta_{\rm c} &= \frac{\pi n_{\rm s} u_{\rm s}^{2}}{M} 
                ( \overline{S^{2}_{\perp}} - \overline{S^{2}_{z}} ) \, \nu_{-}  . 
\nonumber
\end{align}
 Thus Eq.~(\ref{result_jel_micro_grad}) consists of
a term proportional to $\sigma_{\uparrow} - \sigma_{\downarrow}$ 
and a term proportional to $\sigma_{\uparrow} + \sigma_{\downarrow}$. 
 This suggests the spin-dependent force that magnons exert on  electrons,  
$ {\bm F}_\sigma = \sigma  {\bm F}_{\rm s} + {\bm F}_{\rm c}$, contains a 
spin-independent part ${\bm F}_{\rm c}$ as well, where 
\begin{align}
 {\bm F}_{\rm s}  
&= - \frac{1}{2 s_{0}} 
 \left(  1 - \frac{\beta_{\rm s}}{\alpha} \right) {\bm \nabla} \Omega_{\rm mag} , 
\nonumber
\\
 {\bm F}_{\rm c}  
&= - \frac{1}{2 s_{0}} 
 \left( - \frac{\beta_{\rm c}}{\alpha} \right) {\bm \nabla} \Omega_{\rm mag} . 
\nonumber
\end{align}
 Similarly, the  spin-dependent nonequilibrium chemical potential,  
$\delta \mu_\sigma = \sigma \delta \mu_{\rm s} + \delta \mu_{\rm c}$, 
contains a spin-independent part $\delta \mu_{\rm c}$, where 
\begin{align}
 \delta \mu_{\rm s}
&= \frac{1}{2 s_{0}} 
	\left(  1 - \frac{\beta_{\rm s}}{\alpha} \right) \delta \Omega_{\rm mag} , 
\nonumber
\\
 \delta \mu_{\rm c}
&= \frac{1}{2 s_{0}} 
	\left( - \frac{\beta_{\rm c}}{\alpha} \right) \delta \Omega_{\rm mag} . 
\nonumber
\end{align}
 The presence of the charge current (the term proportional to $\beta_{\rm c}$) 
seems interesting from the viewpoint of the momentum transfer effect. 
 However, the sign of $\beta_{\rm c}$ here is not definite and it seems difficult to give   
a clear physical meaning. 
 It vanishes when the magnetic impurities are isotropic, $\overline{S^{2}_{\perp}} = \overline{S^{2}_{z}}$, anyway.  
 Hence we will not pursue this aspect in this paper.
\bibitem{com_deltaT} If a comparison is made with Eq.~(\ref{def_deltaT}), 
we see that $\delta T = 0$, namely, the electron temperature is not changed 
by the coupling to nonequilibrium magnons.
\end{thebibliography}
\end{document}